\documentclass[aps,pra,superscriptaddress]{revtex4-2}

\usepackage{graphicx}
\usepackage{amsmath,amssymb,amsfonts}
\usepackage{bm}
\usepackage{hyperref}

\begin{document}

\renewcommand{\thesection}{\arabic{section}}
\renewcommand{\thesubsection}{\arabic{section}.\arabic{subsection}}

\makeatletter
\renewcommand{\p@subsection}{}
\renewcommand{\p@subsubsection}{}
\makeatother

\title{Limits to the Hall effect and other nonreciprocal effects\\ in three-dimensional metamaterials}

\author{Christian Kern}
\email{physics@chrkern.de}
\affiliation{Department of Civil and Mechanical Engineering, Technical University of Denmark, 2800 Kgs. Lyngby, Denmark}
\author{Graeme W. Milton}
\email{graeme.milton@utah.edu}
\affiliation{Department of Mathematics, University of Utah, Salt Lake City, UT 84112, USA}

\begin{abstract}
Metamaterials can exhibit exotic nonreciprocal properties, yet corresponding fundamental limits and design blueprints achieving them are largely unexplored. Here, we derive comprehensive bounds on the effective nonreciprocal properties of three-dimensional metamaterials and identify microstructures achieving or approaching these bounds. We assume that the underlying equations are equivalent to the conductivity problem in a weak applied magnetic field. While we focus on the Hall effect, our results are more generally applicable, particularly to the Faraday effect in the quasistatic regime and in the absence of losses and resonances. Our bounds yield three important implications: First, the effective Hall mobility of a metamaterial cannot be larger than the largest Hall mobility among the constituent materials. Second, under additional conditions, the effective Verdet constant cannot be enhanced either. Third, for diagonal Hall tensor components, the optimal values are achieved either by one of the pure phases or a rank-1 laminate formed from them, provided that the Hall coefficients of all phases have the same sign. Our work elucidates the limits of nonreciprocal metamaterials and identifies key prerequisites for obtaining exotic phenomena such as sign-inversions and enhancements. Several extensions appear within reach, for example to the Faraday effect in metamaterials exhibiting plasmonic resonances.
\end{abstract}


\maketitle

\section{Introduction}\label{sec1}

\noindent
Metamaterials, i.e., artificial composite materials with rationally designed microstructures, are well-known to exhibit effective material properties that go far beyond the properties of their constituent materials. In many cases, the effective properties surpass those of any known conventional material or are at least highly exotic. The range of examples is vast, encompassing phenomena such as negative refractive indices \cite{Smith:2000:CMS,Smith:2000:NRI}, near-zero indices \cite{Kinsey:2019:NZI}, artificial magnetism \cite{Schelkunoff:1952:ATP}, auxetic or dilational behavior \cite{Almgren:1985:ITD,Lakes:1987:FSN,Milton:1992:CMP,Sigmund:1995:TMP}, negative and/or anisotropic mass densities \cite{Auriault:1985:DCE,Auriault:1994:AHM,Zhikov:2000:EAT,Sheng:2003:LRS}, and other extreme elastic responses \cite{Milton:1995:WET,Jaglinski:2007:CMV,Gros:2023:TMP}. With the advance of additive manufacturing technologies, metamaterials offer the prospect of readily obtaining tailor-made material properties from only a small set of constituent materials. Moreover, metamaterials allow for tailored spatial variations of the effective properties, which can be exploited to guide and manipulate waves in unprecedented ways -- for instance, in cloaking applications \cite{Dolin:1961:PCT,Greenleaf:2003:ACC,Alu:2005:ATP,Milton:2006:CEA,Leonhardt:2006:OCM,Pendry:2006:CEM,Milton:2006:CEM}.

\medskip
\noindent
Despite this vast design space, the effective properties of metamaterials are not limitless. Rather, their range -- as the microgeometry is varied -- is bounded as a function of the properties of the constituent materials. Over the past few decades, a plethora of such bounds has been derived covering a vast range of physical effects \cite{Milton:2002:TOC}. Prominent examples include the Hashin-Shtrikman bounds on the effective conductivity \cite{Hashin:1962:VAT} and the effective elastic moduli \cite{Hashin:1963:VAT} of isotropic two-phase metamaterials and the Bergman-Milton bounds \cite{Bergman:1980:ESM,Milton:1980:BCD, Milton:1981:BCP, Milton:1981:BTO,Bergman:1982:RBC} on (generally) complex-valued effective material parameters of two-phase metamaterials.

\begin{figure}[h!]
	\makebox[\textwidth]{\includegraphics{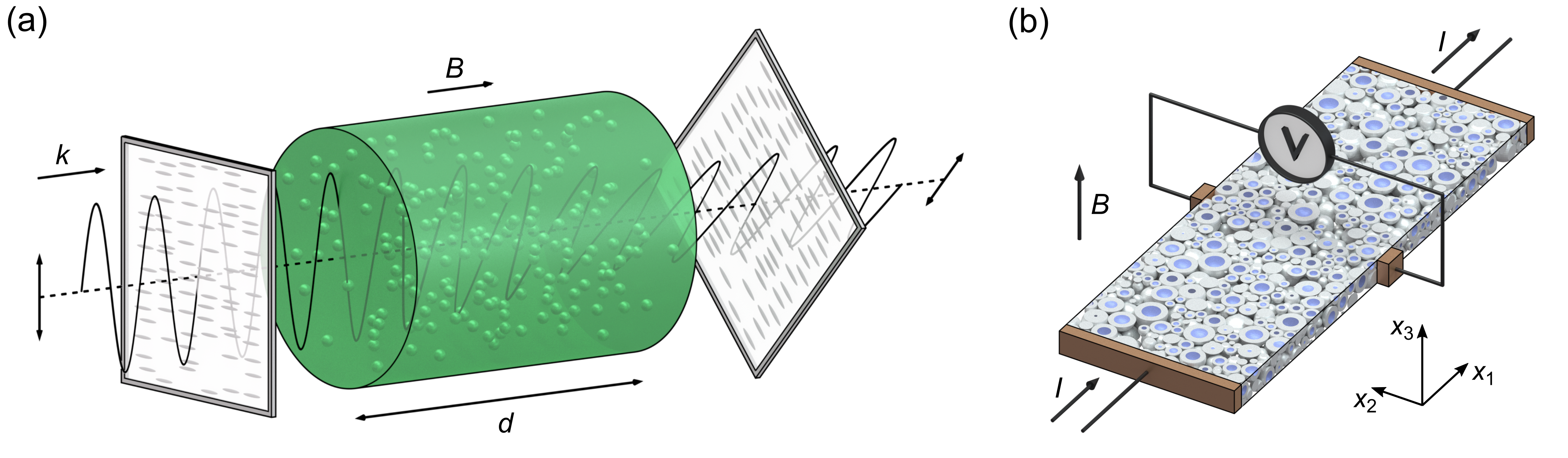}}
	\caption{Schematic illustration of two nonreciprocal metamaterial devices. (a) Optical isolator made from a magneto-active medium and two polarizers. The magneto-active medium contains small metal particles leading to plasmonic resonances thereby enhancing the Faraday rotation. (b) Hall bar made from a coated sphere assemblage. The phase used for the shell is highly conducting and has nonzero Hall coefficient whereas the core phase is weakly conducting. If the shells become thin, the effective Hall coefficient and thus the Hall voltage becomes anomalously large.}
	\label{fig1}
\end{figure}

\medskip
\noindent
Among the many physical phenomena studied in metamaterials, nonreciprocal effects have seen particular interest in the past few years \cite{Coulais:2017:SNR,Caloz:2018:ENR,Nassar:2020:NRA}. Nonreciprocity is the absence of symmetry with respect to the interchange of sources and receivers in a physical system. Mathematically, nonreciprocity is reflected in an antisymmetric contribution to the tensor describing the material properties. The canonical textbook example is the Faraday effect, which describes the rotation of the plane of polarization of an electromagnetic wave propagating through a magneto-active medium in the presence of a magnetic field. Notably, a Faraday rotator can be used to construct an optical isolator by combining it with two polarizers, as illustrated \textbf{Figure\,\ref{fig1}(a)}. Another well-known example is the Hall effect \cite{Hall:1879:MEC}, which describes, in its simplest form, the appearance of a voltage (the so-called Hall voltage) in a current-carrying slab of material that is placed within a magnetic field, as illustrated in Figure\,\ref{fig1}(b). While the Hall effect is generally well known and has many applications, its nonreciprocal nature appears to be sometimes overlooked. This may be attributed to the fact that, although the Hall effect was initially considered for nonreciprocal devices \cite{Mason:1953:HEG}, most current implementations are not based on it \cite{Wick:1954:FGG,Viola:2014:HEG}.

\smallskip

\noindent
The recent interest in nonreciprocal effects is partly motivated by the large practical relevance of nonreciprocal devices, such as optical isolators, and the weakness of magnetooptic effects, which causes the devices to be bulky. Thus, magnetless approaches \cite{Sounas:2017:NRT} and techniques of enhancing magnetooptic effects \cite{Inoue:1998:MOP,Jain:2009:PEF,Temnov:2010:MPM,Chin:2013:NRF,Christofi:2018:GEF,Tyboroski:2023:ENR} are being pursued. For example, already in 1987, it was shown theoretically that the effective Verdet constant of a dilute suspension of small metal particles can become anomalously large \cite{Hui:1987}. Another avenue is to reconsider the Hall effect for the realization of nonreciprocal devices \cite{Viola:2014:HEG,Mahoney:2017:QHC}. Furthermore, in the case of the Hall effect, exotic properties such as the sign-inversion of the effective Hall coefficient, which contradicts common textbook wisdom, have attracted considerable attention \cite{Briane:2009:HTD,Kadic:2015:HES,Kern:2016:EES}. Under certain conditions, the effective Hall coefficient can become anomalously large \cite{Briane:2009:GHE}. In particular, it has been shown that the effective Hall coefficient diverges at the percolation threshold \cite{Bergman:1983:HPT,Rohde:1987:HPT,Dai:1987:HPS}. Furthermore, in metamaterials the effective Hall electric field can be parallel, rather than perpendicular, to the magnetic field \cite{Briane:2010:AEH,Kern:2015:PHE,Kern:2017:EPH}. In the high-field regime, the magnetoresistance can be strongly anisotropic, even in microstructures with a high-degree of symmetry \cite{Bergman:1994:CSF,Tornow:1996:AMC,Bergman:1997:RAM}.
\smallskip

\noindent
Surprisingly, despite this large interest in the subject and the high practical relevance, limits to nonreciprocal effects in three-dimensional metamaterials have been rarely studied. Rather, much of the work has considered two-dimensional or columnar microstructures \cite{Dykhne:1970:APR,Shklovskii:1976:CBH,Stroud:1984:NER,Briane:2008:HTD}. For two-dimensional two-phase metamaterials, Milton fully characterized the set of realizable effective conductivity tensors, even for large magnetic fields \cite{Milton:1988:CHE}. Exceptions in which the full three-dimensional case was considered include the work of Briane and Milton, who derived bounds for the strong-field regime \cite{Briane:2011:BSF} using the Cherkaev-Gibiansky-Milton transformation \cite{Cherkaev:1994:VPC,Milton:1991:FER} and for the weak-field regime via a perturbation approach \cite{Briane:2009:GHE}. Using a minor modification to the derivation of the latter bounds, it was later shown that the effective Hall mobility cannot be larger than twice the largest Hall mobility among the phases \cite{Kern:2018:THE}. While these bounds have significantly advanced our understanding, they are still far from being tight.
\medskip

\noindent
In this paper, we address this open problem. After introducing the mathematical formalism and the properties of interest (\textbf{Section 2}), we will use a monotonicity argument to reduce the problem of bounding the effective properties of a \textbf{non}reciprocal metamaterial to one of bounding the effective properties of a corresponding fictitious reciprocal metamaterial (\textbf{Section 3}). This will allow us to derive multiphase bounds that vastly improve on the state-of-the-art (\textbf{Section 4}). On the basis of these bounds, we will draw major conclusions on some of the key properties of interest: We will show that neither the effective Hall mobility nor (under certain additional conditions) the effective Verdet constant of a metamaterial can improve on the phases from which it is made. Moreover, we will show that the largest value of any diagonal component of the effective Hall tensor is attained by either one of the pure phases or a rank-$1$ laminate formed from two of the pure phases. Further improvements will be obtained in the case of two-phase metamaterials (\textbf{Section 5}), in which we additionally make use of the field equation recursion method. For two-phase metamaterials we will also obtain bounds that take the volume fraction of the phases into account. Such bounds can be also used in an inverse sense to determine the volume fraction of the phases of a metamaterial sample from a measurement of its effective properties, see, e.g., \cite{Phan-Thien:1982:PUB,McPhedran:1982:ESI,McPhedran:1990:ITP,Cherkaev:1998:IBM}.
\medskip

\newpage
\noindent
Throughout this paper, we will assume that the magnetic field and, thus, the antisymmetric contribution to the conductivity tensor is small. While we focus on the Hall effect, our results apply to other effects as well, as long as they are equivalent in their mathematical description. In particular, we will consider the Faraday effect, in which case our bounds apply in the quasistatic regime. Furthermore, we assume that the tensor describing the material properties is self-adjoint and positive definite, which for the Faraday effect implies the absence of absorption and resonance.

\section{Preliminaries}\label{sec2}

\noindent
The following section briefly introduces the underlying mathematical equations and the effective tensors describing the macroscopic behavior of the nonreciprocal metamaterial. Details can be found in Refs.\,\cite{Kern:2018:THE,Kern:2023:SRE}. For a mathematically rigorous treatment, based on the notion of $H$-convergence, we refer the reader to \cite{Briane:2009:HTD}.
\medskip

\noindent
The equations we are considering take the form
\begin{equation}
	\nabla\cdot\widetilde{\bm{j}}=0, ~\nabla\times\widetilde{\bm{e}}=0, ~\widetilde{\bm{j}}=\widetilde{\bm{\sigma}}(\bm{B})\widetilde{\bm{e}}.
	\label{eq:conductivityprob}
\end{equation}
The first two equations are the differential constraints on the electric current density, $\widetilde{\bm{j}}$, and the electric field, $\widetilde{\bm{e}}$. The third equation is the constitutive law, the microscopic version of Ohm's law, wherein $\widetilde{\bm{\sigma}}(\bm{B})$ is the magnetic-field dependent conductivity tensor and $\bm{B}$ is the magnetic induction. Note that we assume that the magnetic permeability is constant. The tilde indicates that these are the magnetic-field dependent quantities and not their zero magnetic-field (zmf) counterparts.
\medskip

\noindent
Since we assume that the magnetic field is small, it suffices to consider terms up to the first order in the magnetic field. Thus, the constitutive law takes the form \cite{LandauCont,Briane:2009:HTD},
\begin{equation}
	\widetilde{\bm{j}}=\bm{\sigma}\widetilde{\bm{e}}+(\bm{S}\bm{B})\times\widetilde{\bm{e}},
	\label{eq:constrel}
\end{equation}
where $\bm{\sigma}$ is the zmf conductivity tensor and $\bm{S}$ is a rank-two tensor describing the effect of the magnetic field. Depending on the boundary conditions, one may want to work with the quantities appearing in the inverted form of the constitutive law, 
\begin{equation}
	\widetilde{\bm{e}}=\bm{\rho}\widetilde{\bm{j}}+(\bm{A}\bm{B})\times\widetilde{\bm{j}},
\end{equation}
where $\bm{\rho}=\bm{\sigma}^{-1}$ is the zmf\ resistivity tensor and $\bm{A}$ is a rank-two tensor, which generalizes the Hall coefficient, $A$, to anisotropic materials. Thus, for an isotropic material, we have $\bm{A}=A\bm{I}$. The $S$-tensor and the Hall-tensor, $\bm{A}$, are related via, see Prop.\,3 in Ref.\,\cite{Briane:2009:HTD},
\begin{equation}
	\bm{S} = -\text{Cof}\left(\bm{\sigma}\right)\bm{A},
	\label{eq:cof}
\end{equation}
where $\text{Cof}(\cdot)$ is the cofactor matrix.
\medskip

\noindent
The same equations, (\ref{eq:conductivityprob}) and (\ref{eq:constrel}), describe the Faraday effect in magneto-active materials in the quasistatic regime upon making the replacements
\begin{equation}
	\widetilde{\bm{j}} \rightarrow \widetilde{\bm{d}},~\bm{\sigma} \rightarrow \bm{\varepsilon} \text{, and } \bm{S} \rightarrow \frac{\bm{\gamma}}{i\varepsilon_0},
\end{equation}
where $\widetilde{\bm{d}}$ is the electric displacement field, $\bm{\varepsilon}$ is the zero magnetic-field permittivity tensor, which we require to be positive definite, $\bm{\gamma}$ is the magnetogyration tensor, and $\varepsilon_0$ is the vacuum permittivity. Note that our bounds on $\bm{\varepsilon}^*$ and $\bm{\gamma}^*$ imply bounds on the Faraday polarizability, in the same way in which the bounds on the complex permittivity of reciprocal metamaterials in Ref.\,\cite{Kern:2020:TBE} imply bounds on the complex polarizability.
\medskip

\noindent
For problems described by the stationary convection-diffusion equation, see e.g. Ref.\cite{Fannjiang:1994:CED} and \S 53 in Ref.\,\cite{Landau:1987:FM}, nonreciprocity can be caused by a fluid flow with a velocity field $\bm{v}(\bm{x})$. If the flow is incompressible, then the convection-diffusion equation be written as
\begin{equation}
	\nabla\cdot(\bm{D}(\bm{x})\nabla T(\bm{x}))-\bm{v}(\bm{x})\cdot\nabla T(\bm{x})=0,
\end{equation}
where $T(\bm{x})$ is the temperature and $\bm{D}(\bm{x})$ is the diffusivity tensor. (Note that this corrects a sign error in Ref.\cite{Fannjiang:1994:CED}.) As the flow is incompressible, there exists an antisymmetric matrix potential $\bm{H}(\bm{x})$, in terms of which $\bm{v}(\bm{x})=\nabla\cdot\bm{H}(\bm{x})$. Using this potential, the convection-diffusion equation can be written as
\begin{equation}
	\nabla\cdot\left[(\bm{D}(\bm{x})-\bm{H}(\bm{x}))\nabla T(\bm{x})\right]=0.
\end{equation}
Thus, Equation (\ref{eq:conductivityprob}) is equivalent to the stationary convection-diffusion equation for an incompressible flow upon making the replacements
\begin{equation}
	\widetilde{\bm{e}} \rightarrow \nabla T \text{ and } \widetilde{\bm{\sigma}} \rightarrow \bm{D}-\bm{H}.
\end{equation}
Note that this means that our analysis applies when $\bm{H}$ is small, i.e., for small velocities.
\medskip

\noindent
For the rest of the derivation, we return to the notation that is commonly used for the Hall effect, keeping in mind that we can always make the replacements above. We are interested in the corresponding effective tensors $\bm{\sigma}^*$ and $\bm{S}^*$, which describe the behavior of the metamaterial on the macroscopic length-scale,
\begin{equation}
	\langle\widetilde{\bm{j}}\rangle=\bm{\sigma}^*\langle\widetilde{\bm{e}}\rangle+(\bm{S}^*\bm{B})\times\langle\widetilde{\bm{e}}\rangle,
\end{equation}
where $\langle\cdot\rangle$ denotes the average over the unit cell of periodicity.
The corresponding inverted quantities are the effective zmf\ resistivity tensor $\bm{\rho}^*=(\bm{\sigma}^*)^{-1}$ and the effective Hall tensor, $\bm{A}^*$, the latter being linked to the effective $S$-tensor via the macroscopic version of (\ref{eq:cof}).
\medskip

\noindent
The effective zmf\ conductivity tensor can be obtained by solving the zero-magnetic conductivity problem for pairs of fields $\bm{e}$, $\bm{j}$ and subsequently solving the macroscopic zmf\ constitutive law,
\begin{equation}
	\langle\bm{j}\rangle=\bm{\sigma}^*\langle\bm{e}\rangle
	\label{eq:constrel_mac}
\end{equation}
for the components of the zmf\ conductivity tensor. Since we are considering three-dimensional metamaterials, we require three pairs of fields, $\bm{e}=\bm{e}_{(i)}$, $\bm{j}=\bm{\sigma}\bm{e}_{(i)}$ with $i\in\lbrace 1,2,3\rbrace$, to obtain all components of $\bm{\sigma}^*$. One can find elegant expressions for the effective tensors if one requires that $\langle\bm{e}_{(i)}\rangle=\hat{\bm{x}}_i$ and recasts the fields as the columns of a matrix $\bm{E}$, which is commonly referred to as the ''matrix-valued electric field`` \cite{Kern:2018:THE} or the ''corrector matrix`` \cite{Briane:2004:CSC, Murat:1997:C}. Then $\langle\bm{E}\rangle=\bm{I}$, and the effective zmf\ conductivity tensor is given by
\begin{equation}
	\bm{\sigma}^*=\langle\bm{\sigma}\bm{E}\rangle.
\end{equation}

\noindent
The effective $S$-tensor too can be obtained from the matrix-valued field (for zero magnetic field). Using a perturbation approach, it can be shown that \cite{Briane:2009:HTD}
\begin{equation}
	\bm{S}^{*}=\langle\text{Cof}(\bm{E})^{\intercal}\bm{S}\rangle.
	\label{eq:effective_S}
\end{equation}
Note that a precursor of this result was obtained by Bergman \cite{Bergman:1983:SDL}.
\medskip

\noindent
A key goal of this paper is to derive bounds on $\bm{\sigma}^*$ and $\bm{S}^*$ given $\bm{\sigma}$ and $\bm{S}$ and to identify microstructures that attain them or come close to doing so. Instead of trying to determine the full set of attainable properties (the so-called $G$-closure), which is a $12$-dimensional problem, we will focus on those quantities and components that are particularly relevant in practice, which we will discuss in the next section. Generally, we will first derive the bounds in terms of components of $\bm{\sigma}^*$ and $\bm{S}^*$ and subsequently discuss the implications for the properties that are of particular interest.
\medskip

\noindent
Throughout the paper, we will assume that the phases from which the metamaterial is formed are isotropic. In the small magnetic-field limit, their magnetic-field dependent conductivity tensors can therefore be written as   
\begin{equation}
	\widetilde{\bm{\sigma}}=\begin{pmatrix} \sigma_0 && -\eta_i && 0 \\ \eta_i && \sigma_0 && 0 \\ 0 && 0 && \sigma_0 \end{pmatrix} \text{ with } \eta_i = S_iB,
\end{equation}
where we have assumed that the magnetic field is aligned with the $x_3$-direction. Furthermore, we will initially assume that the metamaterial is uniaxial (in the sense that its effective tensors are uniaxial). Thus, the effective magnetic-field dependent conductivity tensor takes the form
\begin{equation}
	\widetilde{\bm{\sigma}}^*=\begin{pmatrix} \sigma_{\perp}^* && -\eta^* && 0 \\ \eta^* && \sigma_{\perp} && 0 \\ 0 && 0 && \sigma_{\parallel} \end{pmatrix} \text{ with } \eta^* = S_{\parallel}^*B,
\end{equation}
provided that the metamaterial is aligned with the magnetic field, i.e., its symmetry axis is parallel to the magnetic field. From the uniaxial bounds, we will subsequently draw conclusions for fully anisotropic metamaterials.

\subsection{Material properties of interest}\label{secproperties}

\noindent
Hall effect experiments and devices often use Hall-plate like geometries with a perpendicular magnetic field \cite{Popovic:2003:HED}. A corresponding example is illustrated in Figure\,1(a). The measured quantity is usually the Hall voltage, which, in terms of the current, $I$, is given by 
\begin{equation}
	U_{\text{H}} = G_{\text{H}}A\frac{1}{l_3}IB_3,
	\label{eq:HallVoltageCurrent}
\end{equation}
where $I$ is the electric current, $l_3$ is the thickness of the Hall plate, and $G_{\text{H}}$ is a geometry factor depending on the (macroscopic) shape (with $G_{\text{H}}\rightarrow 1$ for a long Hall bar) that can be calculated from the response of a homogeneous Hall plate. Thus, the key material property determining the Hall voltage for a prescribed current flow is the Hall coefficient, $A=-S/\sigma^2$. Unsurprisingly, much of the previous work on Hall effect metamaterials considered the effective Hall coefficient, in particular demonstrating giant enhancements \cite{Briane:2009:GHE} and sign-inversions \cite{Briane:2009:HTD,Kern:2016:EES,Kern:2023:SRE}.
\medskip

\noindent
The Hall coefficient is however not the only relevant quantity. If, instead of prescribing a current, one applies a voltage $U$, across the plate, the equation for the Hall voltage becomes
\begin{equation}
	U_{\text{H}} = G_{\text{H}}\mu\frac{l_2}{l_1}UB_3,
\end{equation}
where $l_1$ and $l_2$ are the length and width of the Hall bar, respectively, and $\mu=\sigma A=-S/\sigma$ is the Hall mobility. Furthermore, it can be seen easily that the ratio of the Hall voltage to the offset voltage, the latter being caused by asymmetrically placed sense contacts or asymmetries within the sample, is determined by the Hall mobility too. 
\medskip

\noindent
This motivates us to additionally look for bounds on the effective Hall mobility. In particular, we will be able to conclude that, under the assumptions underlying our derivation, the largest value of the absolute effective Hall mobility is always attained by one of the pure phases from which the metamaterial is formed. 
\medskip

\noindent
As regards the Hall coefficient, we will broaden the discussion to anisotropic materials, which are described by a Hall tensor, $\bm{A}$, instead of a Hall coefficient, $A$. In fact, a Hall bar may just as well be made from an anisotropic material. In such an instance, the diagonal Hall tensor component corresponding to the magnetic field direction (assuming that the coordinate system is chosen such one of its axes aligns with the magnetic field), here $A_{33}$, plays the role of the Hall coefficient in Equation\,(\ref{eq:HallVoltageCurrent}).
\medskip 

\noindent
To obtain bounds on the diagonal components of a (generally) anisotropic effective Hall tensor, we will follow a two step strategy: First, we will derive bounds applying to uniaxial metamaterials. Specifically, we will look for bounds relating the axial component of the effective $S$-tensor, $S_{\parallel}^*$ to the transverse component of the effective conductivity tensor, $\sigma_{\perp}^*$. This immediately implies bounds on the axial component of the effective Hall tensor, $A_{\parallel}^*$. Second, we will use our uniaxial bounds to derive bounds applying to arbitrary anisotropic metamaterials. Our argument uses the idea of forming a fictitious higher symmetry metamaterial from the original anisotropic one. Applying bounds to the former then implies bounds on the latter \cite{}. To obtain this higher symmetry metamaterial, we introduce a specific lamination scheme using two lamination steps (in which the structure is laminated with mirror copies of itself) before, in the last step, forming a polycrystal from the resulting rank-$2$ laminate. Corresponding details can be found in the Supporting Information. We obtain that our uniaxial bounds apply to anisotropic metamaterials upon making the replacements
\begin{equation}
	\sigma_{\perp}^* \rightarrow \sqrt{\frac{\text{det}\left(\bm{\sigma}^{*}\right)}{\sigma_{33}^{*}}} \text{ and } A_{\parallel}^{*} \rightarrow A_{33}^{*},
	\label{eq:poStens}
\end{equation}
where $\bm{\sigma}^{*}$ and $\bm{A}^{*}$ are the effective conductivity and the effective Hall tensor of the anisotropic metamaterial. Note that we are free to choose the coordinate axes and, in this way, obtain a continuum of bounds, even in the uniaxial case.
\medskip

\noindent
This argument will allow us to make another fundamental statement: The largest (absolute) value of any diagonal component of the effective Hall tensor is achieved either by a pure phase or by a rank-$1$ laminate formed from two of the pure phases.
\medskip

\noindent
Furthermore, introducing another, related microstructure scheme (see the Supporting Information), again including two lamination steps and the formation of a polycrystal, we can show that the uniaxial bounds also apply to fully anisotropic metamaterials upon making the alternative replacement 
\begin{equation}
	\sigma_{\perp}^{*} \rightarrow \sqrt{\sigma_{11}^{*}\sigma_{22}^{*}-\left.\sigma_{12}^{*}\right.^2}=\sqrt{\text{Cof}\left(\bm{\sigma}^{*}\right)_{33}} \text{ and } S_{\parallel}^{*} \rightarrow S_{33}^{*}.
	\label{eq:faeq}
\end{equation}
The $S$-coefficient and, in the anisotropic case, the diagonal components of the $S$-tensor determine the sign and magnitude of the current flow induced by the magnetic field in a short Hall bar.
\medskip

\noindent
The isotropic case, for which we hypothesize that the largest value of the effective Hall coefficient is either attained by one of the phases or by a coated sphere assemblage (again depending on the properties of the phases), can be tackled with the same methods and will be further discussed elsewhere \cite{Kern:2025:MBH}. 
\medskip

\noindent
Additionally, we will consider the Faraday effect. A sketch showing a typical configuration for an optical isolator based on a Faraday rotator is shown in Figure\,1(a). The angle of rotation $\alpha$ of the plane of polarization is determined by a material parameter called the Verdet constant, $\mathcal{V}$, 
\begin{equation}
	\alpha = \mathcal{V}Bd \text{ with } \mathcal{V}=-\frac{\pi}{\lambda_0}\frac{\gamma}{\sqrt{\varepsilon}},
	\label{eq:Verdet_constant}
\end{equation}
where $d$ is the thickness of the magneto-optic material along the optical axis, $\lambda_0$ is the vacuum wavelength, $\gamma$ is the magnetogyration coefficient, and $\varepsilon$ is the relative permittivity. The properties of the magneto-optic material need not be isotropic. Rather, the material is typically uniaxial with the electromagnetic wave propagating along the optic axis to which the static magnetic field is collinear. As the mathematical descriptions are equivalent (in the quasistatic regime and for small static magnetic fields), we can make the identification  
\begin{equation}
	\mathcal{V}_{\parallel} \leftrightarrow -\frac{\pi i \varepsilon_0}{\lambda_0}\frac{S_{\parallel}}{\sqrt{\sigma_{\perp}}}.
\end{equation}
\medskip

\noindent
Using this equivalence, we will be able to conclude that, in the two-phase case, the absolute value of the effective Verdet constant cannot exceed the largest absolute Verdet constant of the phases. Under additional assumptions, this conclusion also holds for an arbitrary number of phases.
\medskip

\noindent
While we consider the discussed quantities and symmetries to be of particular importance, it is clear that bounds on other (or relating other) components are relevant as well. One example may be the off-diagonal components of the effective Hall tensor which give rise to a Hall voltage parallel to the magnetic field \cite{Briane:2010:AEH,Kern:2015:PHE,Kern:2017:EPH}. It is to be expected that our approach would still yield useful results in such cases. Ultimately, one should strive to characterize the entire $G$-closure. While this is certainly an ambitious goal, we consider the present paper to constitute a first step in this direction.

\subsection{Translations}

\noindent
To obtain tight bounds, we will make use of translations. In the simplest case, a translation is the addition of a constant offset to the microscopic conductivity tensor if it causes the effective conductivity tensor to shift by the same offset. More precisely, we will use that if we shift $\widetilde{\bm{\sigma}}(\bm{x})$ by the tensor $-\lambda\bm{T}$, where
\begin{equation}
\bm{T}=\begin{pmatrix} 0 && -1 && 0 \\ 1 && 0 && 0 \\ 0 && 0 && 0 \end{pmatrix},
\end{equation}
or any other constant antisymmetric tensor, then the effective tensor shifts in the same way \cite{Stroud:1984:NER}: 
\begin{equation}
	\widetilde{\bm{\sigma}}^*(\widetilde{\bm{\sigma}}(\bm{x})-\lambda\bm{T})=  \widetilde{\bm{\sigma}}^*(\widetilde{\bm{\sigma}}(\bm{x}))-\lambda\bm{T}.
	\label{eq:trans}
\end{equation}
So if we originally had an $n$-phase metamaterial with the phases having conductivity tensors $\widetilde{\bm{\sigma}}_i$ and effective tensor $\widetilde{\bm{\sigma}}^*$, then by this translation we obtain a metamaterial with the phases having conductivity tensors $\widetilde{\bm{\sigma}}_i-\lambda\bm{T}$ and effective tensor $\widetilde{\bm{\sigma}}^*-\lambda\bm{T}$. Note that we require that the antisymmetric part of the translated tensors of the phases is small. Also note that we are using the term translation in its narrow sense \cite{Milton:2002:TOC}.
\medskip

\noindent
Applying a bound to the translated metamaterial, $\eta^*(\widetilde{\bm{\sigma}}_i-\lambda\bm{T})\leq \mathcal{B}(\widetilde{\bm{\sigma}}_i-\lambda\bm{T})$, where $\mathcal{B}$ denotes the bound, implies the bound $\eta^*(\widetilde{\bm{\sigma}}_i)\leq \mathcal{B}(\widetilde{\bm{\sigma}}_i-\lambda\bm{T})+\lambda$, which is often tighter than the original bound, $\eta^*(\widetilde{\bm{\sigma}}_i)\leq \mathcal{B}(\widetilde{\bm{\sigma}}_i)$ (of course, this also holds for a lower bound). However, in order to obtain a tight bound, the value of the factor $\lambda$ has to be chosen carefully. In many instances, there is a unique choice of $\lambda$ resulting in the tightest bound (among all of those translations). In the discussion of the multi-phase bounds below, we will elaborate on how $\lambda$ should be chosen depending on the properties of the phases. Note that in the multiphase case, we will encounter bounds on the absolute value of $\eta^*$. Hence, the implied bound becomes
\begin{align}
	\eta^*(\widetilde{\bm{\sigma}}_i)&\leq \mathcal{B}(\widetilde{\bm{\sigma}}_i-\lambda\bm{T})+\lambda && \text{if } \eta^*(\widetilde{\bm{\sigma}}_i) \geq \lambda \label{eq:trBounds}\\ \text{and } \eta^*(\widetilde{\bm{\sigma}}_i)&\geq -\mathcal{B}(\widetilde{\bm{\sigma}}_i-\lambda\bm{T})+\lambda && \text{if } \eta^*(\widetilde{\bm{\sigma}}_i) \leq \lambda \nonumber.
\end{align}
For the two-phase bounds, we will, without loss of generality by making a translation if necessary, set $S_2=0$ and label the phases such that $\sigma_1>\sigma_2$. Note that it is not a priori clear that this choice corresponds to an optimal translation. However, we were not able to improve on the two-phase bounds by considering other translations. Thus, for $S_2\neq0$, one should consider the implied bounds $(\ref{eq:trBounds})$. Similarly, the analytic expression of the effective microstructures that we will give assume $S_2=0$. The corresponding expressions for $S_2\neq0$ can be readily obtained using Equation\,(\ref{eq:trans}).

\subsection{Optimized microstructures}

\noindent
In addition to deriving bounds, we will introduce optimal microstructures, i.e., metamaterial design having the most extreme properties possible. Additional microstructures will show near-optimal behavior or at least the most extremal behavior known so far. In particular, we will be discussing specific hierarchical laminates, i.e., microstructures formed via a hierarchical layering process on (ideally) infinitely separated length-scales. We determined the effective zmf\ conductivity of the laminates via the usual lamination equations \cite{Tartar:1985:EFC,Milton:1990:CSP,Zhikov:1991:EHM}. The effective $S$-tensor was found by determining the microscopic electric field, $\bm{E}$, and using the perturbation expression (\ref{eq:effective_S}). For a comprehensive introduction to laminates see Chapter\,9 in Ref.\,\cite{Milton:2002:TOC}. 

\medskip

\noindent
For some of the laminates, we will discuss equivalent microstructures, specifically the Hashin-Shtrikhman coated cylinder assemblages. These structures are known to have the same zmf\ conductivity as their laminate counterparts. We have confirmed that this also holds true for the effective $S$-tensor by performing the corresponding calculations. Note that it is not always necessary to perform the (full) calculations: One of the coated cylinders corresponds to an intersection of the volume-fraction dependent bounds, implying that $S_{\parallel}^*$ is uniquely determined by $\sigma_{\perp}^*$ and $f_1$. Thus, any other microstructure having the same effective transverse conductivity and volume-fraction will also have the same axial $S$-coefficient. Another example beyond the coated cylinder assemblage would be an elongated version of the Vigdergauz structure \cite{Vigdergauz:1994:TDG,Grabovsky:1996:BEM}.

\section{Monotonicity of the Effective Tensor}\label{sec3}

\noindent
A key realization underlying the derivation of our bounds is that we can estimate the problem of bounding the non-symmetric tensor $\widetilde{\bm{\sigma}}^*$ by one of bounding a corresponding real symmetric tensor $\bm{\sigma}^{*+}$. To do so, we will replace the non-symmetric conductivity tensors of the phases $\widetilde{\bm{\sigma}}_i$ by suitably chosen real and symmetric tensors $\bm{\sigma}_i^+$. Assuming that in fact the $\widetilde{\bm{\sigma}}_i$ are Hermitian (see the discussion below), we make use of the monotonicity of the effective tensor, i.e., of the fact that if the conductivity tensor of at least one of the phases increases, the effective conductivity tensor increases as well, see, e.g. Ref.\,\cite{Tartar:1979:ECH}. Thus, we choose  $\bm{\sigma}_i^+$ such that $\bm{\sigma}_i^+\geq\widetilde{\bm{\sigma}}_i$. Then, the variational principle
\begin{equation}
	\bm{e}_0^\dagger\bm{\sigma}^{*+} \bm{e}_0 =  \min_{\begin{matrix}\bm{e} \cr \nabla\times\bm{e}=0 \cr \langle \bm{e}\rangle =\bm{e}_0 \end{matrix}}\langle\bm{e}^\dagger\bm{\sigma}^+ \bm{e}\rangle 
	\geq \min_{\begin{matrix}\bm{e} \cr \nabla\times\bm{e}=0 \cr \langle \bm{e}\rangle =\bm{e}_0 \end{matrix}}\langle \bm{e} ^\dagger\tilde{\bm{\sigma}}\bm{e}\rangle=\bm{e}_0^\dagger\widetilde{\bm{\sigma}}^*\bm{e}_0,
\end{equation}
(where the metamaterial geometry and fields are assumed to be periodic and the angular brackets $\langle\cdot\rangle$ denote an average over the unit cell of periodicity) implies
\begin{equation}
	\bm{\sigma}^{*+} \geq \widetilde{\bm{\sigma}}^*.
	\label{eq:estimatebound}
\end{equation}
Hence, an upper bound on the fictitious real and symmetric effective tensor $\bm{\sigma}^{*+}$ will provide an upper bound on $\tilde{\bm{\sigma}}^*$. The former problem is significantly more tractable.

\medskip

\noindent
As regards our assumption that the $\widetilde{\bm{\sigma}}_i$ are Hermitian: If one considers the Faraday effect, i.e., if the magnetic-field dependent permittivity, $\widetilde{\bm{\varepsilon}}$, replaces the conductivity, this assumption holds if the phases are lossless. In this case, the zmf\ permittivity tensor will be purely real (and symmetric) and the antisymmetric contribution due to the magnetic field will be purely imaginary. In the case of the Hall effect, the antisymmetric contribution is purely real. However, since we are only interested in the effective behavior for weak magnetic fields, which corresponds to merely bounding the derivatives $\partial \widetilde{\bm{\sigma}}^*/\partial \eta_i$ evaluated at $\eta_i=0$, we can, without loss of generality, because $\widetilde{\bm{\sigma}}^*$ is an analytic function, assume that the antisymmetric contribution is purely imaginary. This is equivalent to taking a magnetic field that is purely imaginary. A related route was taken by Schulgasser \cite{Schulgasser:1976:BEP} for bounding the complex dielectric constant of two phase metamaterials to first order in the imaginary parts of the complex dielectric constants of the (low loss) phases. Note that, in the same spirit, we could include small losses into our analysis of the Faraday effect, but we do not pursue this further within the present paper.

\medskip

\noindent
We now return to the main part of our derivation: We choose to increase $\widetilde{\bm{\sigma}}_i$ to the isotropic tensor
\begin{equation}
	\bm{\sigma}_i^+= (\sigma_i+|\eta_i|)\bm{I},
\end{equation}
which just barely meets the requirement of positive-semidefiniteness, as the determinants of the upper-left $2\times 2$-minors of the matrices $\bm{\sigma}_i^+-\widetilde{\bm{\sigma}}_i$ evaluate to zero. Furthermore, we note that we are interested in the case of $\widetilde{\bm{\sigma}}^*$ being uniaxial. Hence, $\bm{\sigma}^{*+}$ is uniaxial and Equation\,(\ref{eq:estimatebound}) becomes
\begin{equation}
	\sigma_{\perp}^{*+} \geq \sigma_{\perp}^*+|\eta^*|.
	\label{eq:monotonicityres}
\end{equation}
We also know that $\widetilde{\bm{\sigma}}^*(\eta_i=0)=\bm{\sigma}^*$. We should thus choose an upper bound on $\sigma_{\perp}^{*+}$ that becomes sharp in this case, i.e., that incorporates the known value $\sigma_{\perp}^{*+}(\eta_i=0)=\sigma_{\perp}^*$. 

\medskip

\noindent
Bounds incorporating known values were studied in the past. The works of Bergman \cite{Bergman:1976:VBS} and Prager \cite{Prager:1969:IVB} have particular relevance for our purposes. Note that we are interested in the uniaxial case and, in the first part of this paper, in metamaterials made from more than two phases. In his paper, using the Hashin-Shtrikhman variational principles, Bergman derived bounds for isotropic two-phase metamaterials. Thus, we will have to derive bounds applying more generally. However, we will still be able to use some of Bergman's ideas. Prager's ideas can similarly be used to derive volume-fraction dependent multi-phase bounds, which we discuss in a separate paper \cite{Kern:2025:MBH}.  

\medskip

\noindent
Finally, we note that we can equivalently decrease $\widetilde{\bm{\sigma}}_i$, which yields
\begin{equation}
	\sigma_{\perp}^{*-} \leq \sigma_{\perp}^*-|\eta^*| \text{ for } \bm{\sigma}_i^-= (\sigma_i-|\eta_i|)\bm{I}.
	\label{eq:monotonicityresdecr}
\end{equation}
This is the approach that allows one to employ a lower bound incorporating known values. 

\section{Multi-phase Bounds}\label{sec4}

\noindent
In this section, we will present the multiphase bounds for uniaxial metamaterials obtained based on the monotonicity argument and discuss their attainability. Following the derivation of these bounds, we will use them to draw the aforementioned conclusions on the effective Hall mobility, the effective Hall tensor, and the effective Verdet constant.

\subsection{Statement of the multiphase bounds}

\noindent
The multiphase bounds are given by a set of straight-line segments and/or parabolic arcs in the $(\sigma_{\perp}^*,\,|S_{\parallel}^*)$-plane, each passing through two of the points $(\sigma_i,\,|S_i|)$ corresponding to the pure phases. Whether one obtains a straight line or a parabolic arc for a given segment depends on the mobilities of the two pure phases. If the more conductive phase has a larger mobility, i.e., if $|S_l|\sigma_k \geq |S_k|\sigma_l$ for $\sigma_l \geq \sigma_k$, then one obtains the straight line passing through the two points, i.e.,
\begin{equation}
	|S_{\parallel}^*|\leq\frac{|S_k|(\sigma_l-\sigma_{\perp}^*)+|S_l|(\sigma_{\perp}^*-\sigma_k)}{\sigma_l-\sigma_k}.
	\label{eq:bound_sl}
\end{equation}
If, contrarily, the more conductive phase has a smaller mobility, i.e., if $|S_l|\sigma_k \leq |S_k|\sigma_l$ for $\sigma_l \geq \sigma_k$, then one obtains a specific parabolic arc passing through the two points, namely
\begin{equation}
	|S_{\parallel}^*|\leq \sigma_{\perp}^*\frac{|S_l|(\sigma_{\perp}^*-\sigma_k)/\sigma_l+|S_k|(\sigma_l-\sigma_{\perp}^*)/\sigma_k}{\sigma_l-\sigma_k}.
	\label{eq:bound_pa}
\end{equation}
In this manner, one obtains either a straight-line segment or a parabola arc for each pair of pure phases. However, not all of these curves are bounds. Rather, the bounds are given by those straight-line segments and parabolic arcs for which all other points $(\sigma_i,\,|S_i|)$ lie below or on the extended line segment/ parabolic arc. The resulting bounds are illustrated in \textbf{Figure\,2}.\\
\begin{figure}[h!]
	\centering
	\includegraphics{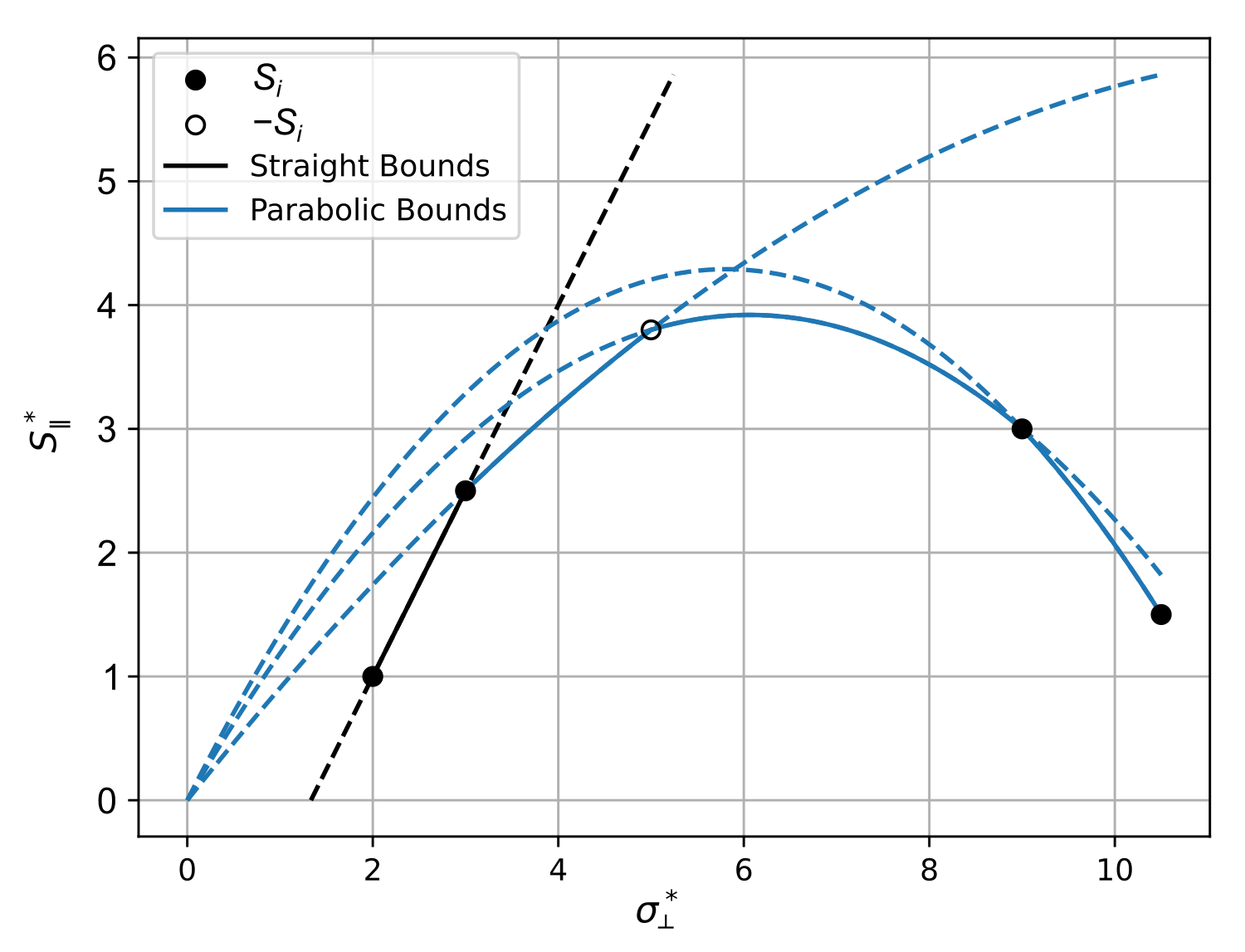}
	\caption{Illustration of the (volume-fraction independent) multi-phase bounds in the $(\sigma_{\perp}^*,\,|S_{\parallel}^*|)$-plane. The bounds are given by a set of straight-line segments (black lines) and parabolic arcs (blue curves), each passing through two of the points $(\sigma_i,\,|S_i|)$. In the depicted case, but not generally, we obtain only a single straight-line segment. All points $(\sigma_i,\,|S_i|)$ corresponding to the pure phases lie below or on the extended line segments/parabolic arcs (dashed curves). Parameters of the five phases are $\sigma=(2,\,3,\,5,\,9,\,10.5)$ and $S=(1,\,2.5,\,-3.8,\,3,\,1.5)$. Note that one of the pure phases has a negative $S$-coefficient.}
	\label{fig2}
\end{figure}
\medskip

\noindent
Depending on the properties of the phases, the bounds can comprise arbitrarily many straight-line segments and/or parabolic arcs. However, no straight-line segment can lie to the right of a parabolic arc.\\
\medskip

\subsection{Derivation of the multiphase bounds}

\noindent
The first step in the derivation of our multiphase bounds on uniaxial nonreciprocal metamaterials is to find corresponding bounds incorporating a pair of known values. Following Bergman \cite{Bergman:1976:VBS}, we start from the Hashin-Shtrikhman variational principles, which we write in the form, cf. Section\,13.5 in Ref.\,\cite{Milton:2002:TOC},
\begin{equation}
	\bm{e}_0 \cdot \bm{\sigma}^{*-}\bm{e}_0 \geq \bm{e}_0 \cdot \sigma_0^{-}\bm{e}_0 + 2\langle \bm{e}_0\cdot\underline{\bm{p}}^{-} \rangle - \langle \underline{\bm{p}}^{-}\cdot\left[(\sigma^{-}-\sigma_0^{-})^{-1}\bm{I}+\bm{\Gamma}_1/\sigma_0^{-}\right]\underline{\bm{p}}^{-}\rangle,
	\label{eq:Hashin}
\end{equation}
which holds for any trial polarization $\underline{\bm{p}}^{-}$, where $\bm{\Gamma}_1$ is the projection operator onto curl-free fields with zero average value. The bound requires that $\sigma_0^{-} \geq 0$ and  $\sigma_i^- \geq \sigma_0^-$. To incorporate the known values, we choose, like Bergman, the trial field as
\begin{equation}
	\underline{\bm{p}}^{-} = a\bm{p} = a\left[\sigma-\sigma_0\right]\bm{e},\,\langle\bm{e}\rangle=\bm{e}_0,
\end{equation}
where $\sigma$, $\sigma_0$, $\bm{p}$ and $\bm{e}$ are the local material properties and fields associated with the problem for which the effective conductivity is known, which in our case is the zero magnetic-field problem. Hence, we have
\begin{equation}
	\langle\bm{p}\rangle = (\bm{\sigma}^*-\sigma_0\bm{I})\bm{e}_0 \text{ and } \bm{e}_0\cdot \bm{\sigma}^*\bm{e}_0=\langle\bm{e}\cdot\sigma\bm{e}\rangle.
\end{equation}
Then, we obtain from (\ref{eq:Hashin})
\begin{equation}
	\bm{e}_0 \cdot (\bm{\sigma}^{*-}-\sigma_0^{-}\bm{I})\bm{e}_0 \geq 2a\bm{e}_0 \cdot (\bm{\sigma}^*-\sigma_0\bm{I})\bm{e}_0 - a^2\left\langle \left[\frac{\left(\sigma-\sigma_0\right)^2}{\sigma^{-}-\sigma_0^{-}}+\frac{\sigma_0^2}{\sigma_0^{-}}\right]\bm{e}\cdot\bm{e} \right\rangle+a^2\left[\frac{\sigma_0}{\sigma_0^{-}}\right]\bm{e}_0\cdot\bm{e}_0.
\end{equation}
We now consider a uniaxial metamaterial. Setting $\bm{e}_0=(1,0,0)^{\intercal}$, we get
\begin{equation}
	\sigma_{\perp}^{*-}-\sigma_0^{-} \geq 2a(\sigma_{\perp}^*-\sigma_0) - a^2\left\langle \left[\frac{\left(\sigma-\sigma_0\right)^2}{\sigma^{-}-\sigma_0^{-}}+\frac{\sigma_0^2}{\sigma_0^{-}}\right]\bm{e}\cdot\bm{e} \right\rangle+a^2\left[\frac{\sigma_0}{\sigma_0^{-}}\right].
	\label{eq:bounduni}
\end{equation}
In contrast to Ref.\,\cite{Bergman:1976:VBS}, we continue a multiphase analysis and, to estimate the average in (\ref{eq:bounduni}), set 
\begin{equation}
	b=\max_i\left[\frac{\left(\sigma_i-\sigma_0\right)^2}{\sigma_i\left(\sigma_i^{-}-\sigma_0^{-}\right)}+\frac{\sigma_0^2}{\sigma_i\sigma_0^{-}}\right]
\end{equation}
which, with the optimal choice of $a$ and after setting $\sigma_0=0$ gives the lower bound
\begin{equation}
	\sigma_{\perp}^{*-}\geq\sigma_0^-+\frac{\sigma_{\perp}^*}{b} \text{ with } b=\max_i\left[\frac{\sigma_i}{\sigma_i^--\sigma_0^-}\right].
\end{equation}
Assuming that this maximum over $i$ is attained by two pure phases, which we label as $k$ and $l$, we obtain 
\begin{equation}
	\sigma_0^{-}=\frac{\sigma_l\sigma_k^--\sigma_k\sigma_l^-}{\sigma_l-\sigma_k}
\end{equation}
and, thus,
\begin{equation}
	\sigma_{\perp}^{*-}\geq\frac{\sigma_k^-(\sigma_l-\sigma_{\perp}^*)+\sigma_l^-(\sigma_{\perp}^*-\sigma_k)}{\sigma_l-\sigma_k}.
\end{equation}
Making use of the monotonicity result for lower bounds (\ref{eq:monotonicityresdecr}), we finally obtain the straight-line bounds (\ref{eq:bound_sl}). Note that the condition $\sigma_0^-\geq 0$ is equivalent to the mobilities being well-ordered, $|S_l|/\sigma_l\geq |S_k|/\sigma_k$. The condition $\sigma_i^-\geq\sigma_0^-$ holds in the limit when the $\eta$-coefficients are infinitesimally small.\\
\medskip

\noindent
To derive the parabolic-arc bound, we replace $\sigma_0^-$ by $\sigma_0^+$ satisfying $\sigma_0^{+} \geq 0$ and $\sigma_i^+ \leq \sigma_0^+$, reverse the sign of all inequalities, and replace the maximums by minimums. Choosing $\sigma_0 \rightarrow \infty$, we obtain the following upper bound
\begin{equation}
	\sigma_{\perp}^{*+} \leq \left(\sigma_0^++\frac{1}{\tilde{b}\sigma_{\perp}^*}\right)^{-1} \text{ with } \tilde{b} = \max_i\left[\frac{1/\sigma_i}{1/\sigma_i^+-1/\sigma_0^+} \right],
\end{equation}
which holds if $\sigma_0^+\geq 0$ and $\sigma_i^+\leq\sigma_0^+$. Again, assuming that this maximum is attained by two pure phases, which we label as k and l, we obtain
\begin{equation}
	1/\sigma_0^{+}=\frac{1/(\sigma_k\sigma_l^+)-1/(\sigma_l\sigma_k^+)}{1/\sigma_k-1/\sigma_l}
\end{equation}
and, thus,
\begin{equation}
	\sigma_{\perp}^{*+} \leq \frac{\sigma_{\perp}^*\sigma_k^+\sigma_l^+(\sigma_l-\sigma_k)}{\sigma_l^+\sigma_k(\sigma_l-\sigma_{\perp}^*)+\sigma_k^+\sigma_l(\sigma_{\perp}^*-\sigma_k)}.
\end{equation}
Making use of the monotonicity result for upper bounds (\ref{eq:monotonicityres}), keeping terms up to first order in the $\eta$-coefficients, we obtain the parabolic-arc bounds (\ref{eq:bound_pa}). Note that the condition $\sigma_0^+\geq 0$ is equivalent to the mobilities being non-well ordered, i.e., $|S_l|/\sigma_l\leq |S_k|/\sigma_k$. The condition $\sigma_i^+\leq\sigma_0^+$ which holds in the limit when the $\eta$-coefficients are infinitesimally small.

\subsection{Tightening the multiphase bounds using translations}

\noindent
Significantly tightened bounds can be obtained using translations. Specifically, we are considering translations of the antisymmetric contribution to the conductivity tensor, $\eta_i \rightarrow \eta_i-\lambda$, $\eta^* \rightarrow \eta^*-\lambda$, as discussed in detail in Section\,2.2. An application of such translations to our bounds is illustrated in \textbf{Figure\,3}. Note that the bounds resulting from the translation confine the effective parameters to a substantially smaller area. In particular, the bounds are no longer symmetric with respect to the $\sigma^*$-axis.

\medskip

\noindent
To obtain the tightest bounds, the translation parameter $\lambda$ has to be chosen carefully. For the bounds considered here, it is typically possible to determine the best choice using elementary considerations, specifically by considering the effects that translations have on the bounds: First, under a translation, the order of the mobilities of the (translated) pure phases can flip. Thus, a straight-line segment can become a parabolic arc and vice versa. Second, a line segment or arc may cease (or start) to be a bound -- as the other points may no longer (or start to) lie below the extended curve. Third, in the case of a parabolic-arc bound, decreasing the absolute values of the $S$-coefficients tightens the bound. There is no such effect for the straight-line segments. Thus, the optimal translation for a parabolic-arc bound is uniquely determined, whereas we obtain a range of optimal translations in the case of a straight-line bound. Fourth, a translation can be used to ensure that the translated $S$-coefficients of all phases have the same sign, which is generally favorable. For an example how these considerations can be used to find the optimal translations, see Figure\,3.

\medskip

\begin{figure}[h!]
	\centering
	\includegraphics{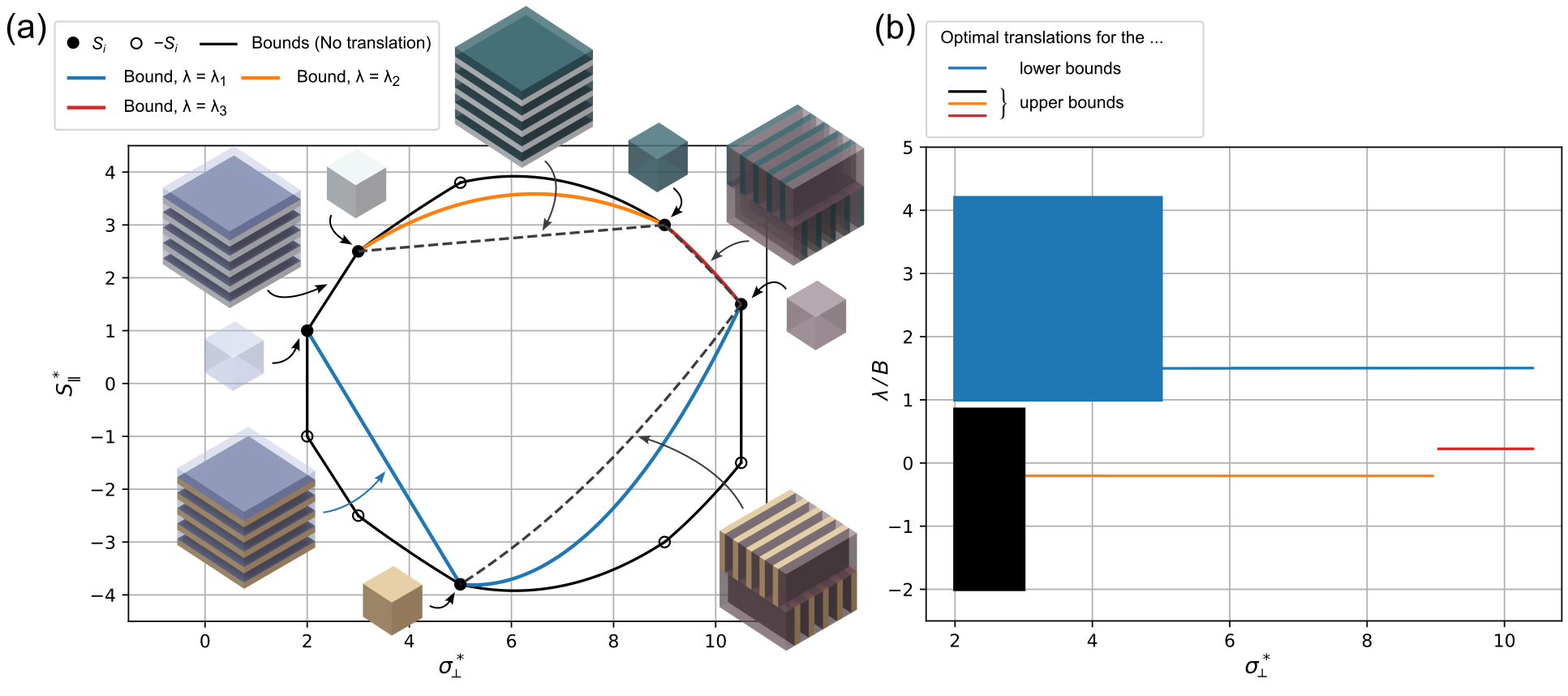}
	\caption{Improvements over the multiphase bounds shown in Figure\,2 obtained via translations. (a) Original bounds (black curves) and improved bounds (colored curves). The black dots correspond to the pure phases. (b) Corresponding values of the translation parameter $\lambda$. Three translations are used: The first translation (blue curves) corresponds to the smallest value of $\lambda$ for which $\eta_3-\lambda$ and $\eta_5-\lambda$ have the same sign. The second translation (orange curve) is such that $|\eta_3-\lambda|$, $|\eta_4-\lambda|$, and $|\eta_5-\lambda|$ lie on the same parabolic arc. The third translation (red curve) shifts the parabolic arc as much as possible downwards without $\eta_4-\lambda$ and $\eta_5-\lambda$ having opposite signs. For the third translation, the improvement is too small to be discernible. Note that for those segments of the bounds that are straight lines, a whole range of translations results in the same bounds. The values of the translation parameter (including the limits of the ranges) follow from the considerations discussed in the main text. For example, the lower limit of the range for the upper (black range) corresponds to the transition to a parabolic arc. In panel (a), two types of microstructures attaining or approaching the bounds are illustrated by the insets. The two straight-line bounds are attained by rank-$1$ laminates. The three parabolic-arc bounds are approached by another rank-$1$ laminate (dashed line) and two uniaxial polycrystals (dashed curves). Each of the microstructures is formed from two of the pure phases.}
	\label{fig3}
\end{figure}

\noindent
The attainability of the improved bounds is as follows: The straight-line bounds connecting two of the points corresponding to the pure phases are optimal. They are attained by rank-$1$ laminates formed from the two respective pure phases. Labeling these phases as $i$ and $j$, such a laminate has effective properties
\begin{equation}
	\sigma_{\perp}^* =  f_i\sigma_i + f_j\sigma_j,~S_{\parallel}^* = f_iS_i + f_jS_j,
	\label{eq:effproplam}
\end{equation}
where $f_i$ and $f_j$ are the volume fractions of pure phases $i$ and $j$, respectively. Note that, as the volume fraction is varied, the effective properties of the laminate traverse the entirety of the straight-line bound.\\
\medskip

\noindent
The situation is more subtle for the parabolic-arc bounds, which can seemingly at most be approached: In some instances, the extremal microstructure appears to be again a rank-$1$ laminate. In other instances, the extremal microstructure appears to be a uniaxial polycrystal formed from a rank-$1$ laminate. Specifically, this polycrystal that is formed from a rank-$1$ laminate in the following way: Assume that the layers of the rank-$1$ laminate are perpendicular to the $x_1$-axis. In addition, consider a rotated copy of this laminate with layers perpendicular to the $x_2$-axis. To form the polycrystal, the rank-$1$ laminate and its rotated version are laminated in equal proportions with layers perpendicular to the $x_3$-axis. The effective properties of the resulting polycrystal are uniaxial with the $x_3$-axis being the extraordinary axis. Note that this way of forming a polycrystal corresponds to the first step of a Schulgasser construction \cite{Schulgasser:1977:BCS} and that this microstructure was discussed by Milton in the context of the effective complex permittivity \cite{Milton:1981:BCP}. For an expression for the effective properties see the section regarding two-phase metamaterials below.
\medskip

\noindent
For large conductivity contrasts, there are microstructures exhibiting an effective behavior beyond the rank-$1$ laminate and the uniaxial polycrystal, in particular microstructures exhibiting sign-inversions (which become enhancements if the $S$-coefficients of the pure phases are interchanged) -- see the corresponding discussion below. Note that in the large-conductivity regime, the gap between the bounds and the most extremal microstructures that we were able to identify becomes somewhat large. Nevertheless, even in this regime, the bounds are sufficiently tight to yield major conclusions for some of the most important material properties, as we will now see.

\subsection{Key implications -- multiphase bounds}

\noindent
Starting from the multiphase bounds (\ref{eq:bound_sl}) and (\ref{eq:bound_pa}), we can draw a number of conclusions for the effective properties that we consider to be of particular relevance in practice (see Section\,\ref{secproperties} for a detailed introduction). Naturally, we can only make these conclusions under the assumptions underlying the derivation of the bounds (see above).\\
\medskip

\noindent
Our first conclusion concerns the effective Hall mobility:\\
\medskip

\noindent
\fbox{\begin{minipage}{\textwidth}
		For a three-dimensional isotropic metamaterial composed of an arbitrary number of isotropic phases, the effective Hall mobility, $\mu^*=-S^*/\sigma^*$, cannot exceed, in absolute value, the largest absolute Hall mobility among the phases, i.e., $|\mu^*|\leq \max\limits_i |\mu_i|$.
\end{minipage}}\\
\medskip

\begin{figure}[h!]
	\centering
	\includegraphics{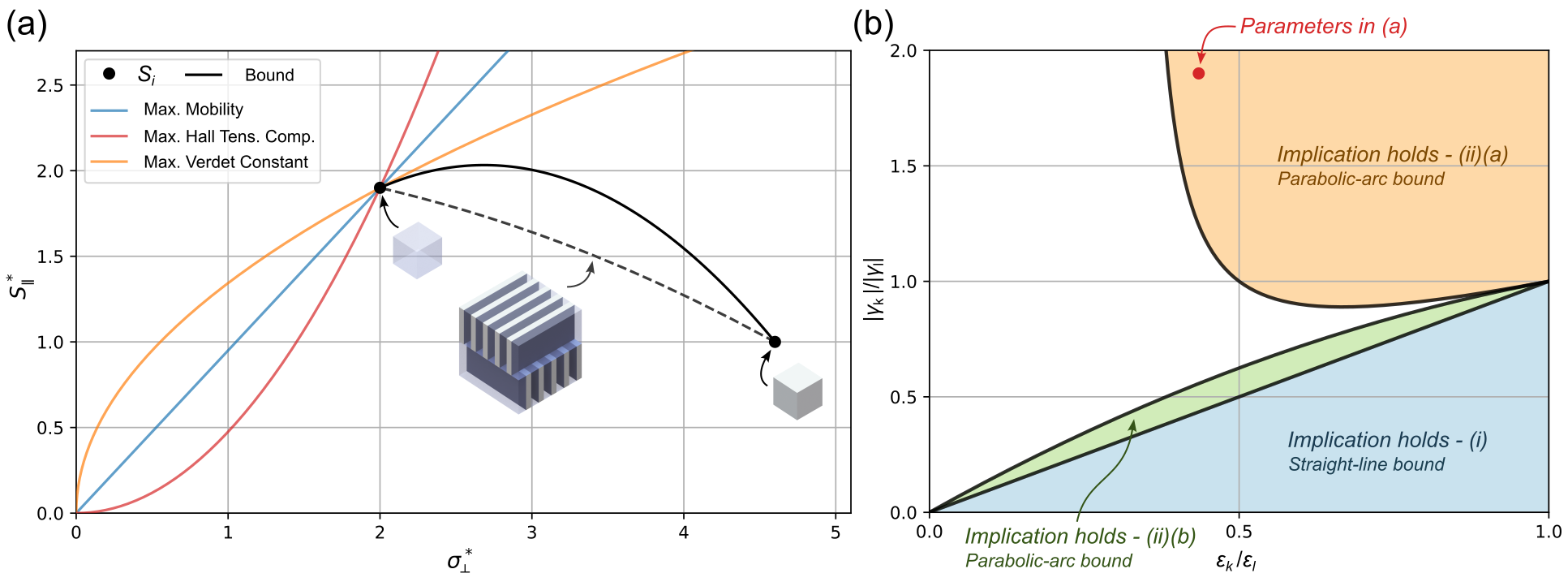}
	\caption{(a) Illustration of a parabolic-arc segment of the multiphase bounds with curves corresponding to the mobility, axial Hall tensor component, and Verdet constant of the left endpoint. For the chosen parameters, the parabolic-arc bound constrains the effective properties to lie below these curves, implying that the maximum is attained by the pure phase corresponding to the left endpoint. The regions corresponding to cases (i), (ii)(a), and (ii)(b) are colored in blue, orange, and green, respectively. The dashed line corresponds to the effective properties of the uniaxial polycrystal described above. (b) Parameter regimes for which the bounds imply that the effective Verdet constant cannot be enhanced. The parameters for (a), $\sigma=(2,\,4.6)$ and $S=(1.9,\,1.0)$, are marked by a red dot. Note that the implication particularly tends to hold in the important regime of small permittivity contrasts (right side of the plot).}
	\label{fig4}
\end{figure}

\noindent
Note that the absolute value of the effective Hall mobility associated with any point in the $(\sigma^*,\,|S^*|)$-plane corresponds to the slope of the straight line passing through the point and the origin. The statement thus holds if the bounds confine the effective properties to lie below the straight line passing through the point corresponding to the phase with the largest mobility and the origin, cf. \textbf{Figure~\ref{fig4}}.\\
\medskip

\noindent
We can draw the first conclusion as follows: Considering each segment of the bounds separately, we can prove the stronger statement that, in the considered segment, the point $(\sigma^*,|S^*|)$ corresponds to a lower Hall mobility than the right endpoint (for a straight-line segment) or the left endpoint (for a parabolic segment). For the straight-line sections, the conclusion follows immediately. For the parabolic-arc segments, we have to show that the parabola does not intersect the straight line passing through the left endpoint and the origin. This follows readily from Equation\,(\ref{eq:bound_pa}).\\
\medskip

\noindent
Our second conclusion concerns the diagonal components of the effective Hall tensor:\\
\medskip

\noindent
\fbox{\begin{minipage}{\textwidth}
		The largest (absolute) value of any diagonal component of the effective Hall tensor, $A_{ii}^*$, of a three-dimensional metamaterial is attained either by one of its pure phases or by a rank-$1$ laminate made from two of its pure phases and with layers perpendicular to the $x_i$-direction, provided the Hall coefficients of all phases have the same sign.
\end{minipage}}
\medskip

\noindent
The second conclusion can be drawn as follows: We first note that it suffices to show that it holds for uniaxial metamaterials, since the result then extends to fully anisotropic metamaterials via the laminate/polycrystal argument discussed in Section\,\ref{secproperties} (see the Supporting Information for details). Thus, we can apply the uniaxial bounds.\\
\medskip

\noindent
We proceed by again considering the segments of the bound separately: For the parabolic segments, we obtain that the largest value of $A_{ii}^*$ corresponds to one of the endpoints and, thus, to a pure phase. For the straight-line segments, the right endpoint (pure phase $l$) yields the largest value of $A_{ii}^*$ when $\sigma_l \leq \sigma_k(s_l+s_k)/(2s_k)$, while the left endpoint (pure phase $k$) is optimal when $\sigma_l \geq 2 \sigma_k s_l/(s_l+s_k)$. Otherwise, the largest value will be attained by the intermediate point on the straight line with the effective transverse conductivity $\sigma_{\perp}^* = 2(s_l \sigma_k - s_k \sigma_l)/(s_l - s_k)$. Under our assumption of same-sign Hall coefficients, $S_k$ and $S_l$ have the same sign as well. Then, the intermediate point corresponds to a rank-1 laminate formed from the pure phases $k$ and $l$ with volume-fraction $f_k=2S_l/(S_l-S_k)+\sigma_l/(\sigma_k-\sigma_l)$ of phase $k$.\\
\medskip

\noindent
We remark that the assumption of same-sign Hall coefficients may be superfluous. However, a corresponding proof would likely require an involved analysis using translations and is beyond the scope of this work.\\
\medskip

\noindent
The mechanism underlying the enhancements can be understood intuitively: At large conductivity contrasts, the current is essentially confined to the highly-conducting layers. A Hall bar made from the rank-$1$ laminate thus effectively acts as if it was thinner. Reducing the thickness of a Hall bar is a strategy that is regularly employed in practice (including by Hall himself \cite{Hall:1879:MEC}).
\medskip

\noindent
Our third conclusion concerns the effective Verdet constant:
\medskip

\noindent
\fbox{\begin{minipage}{\textwidth}
		The absolute value of the effective Verdet constant, $\mathcal{V}_{\parallel}^*$, of a three-dimensional uniaxial metamaterial made from an arbitrary number of isotropic phases cannot be larger than the largest absolute value of the Verdet constant among the phases, i.e., $|\mathcal{V}_{\parallel}^*|\leq \max\limits_i |\mathcal{V}_i|$ if for all segments of the bound one of the following conditions is satisfied:
		\begin{itemize}
			\item[(i)] $|\gamma_l|\varepsilon_k>|\gamma_k|\varepsilon_l$ (i.e., the bound segment is a straight line); or
			\item[(ii)] $|\gamma_l|\varepsilon_k\leq|\gamma_k|\varepsilon_l$ (i.e., the bound segment is a parabolic arc) and
			\begin{itemize}
				\item[(a)] $\varepsilon_k/\varepsilon_l > 3$ and $|\gamma_k|/|\gamma_l| \geq \left[2(\varepsilon_k/\varepsilon_l)^2\right]/\left[3(\varepsilon_k/\varepsilon_l)-1\right]$ or
				\item[(b)] $|\gamma_k|/|\gamma_l| \geq \left[3(\varepsilon_k/\varepsilon_l)-(\varepsilon_k/\varepsilon_l)^2\right]/2$.
			\end{itemize}
		\end{itemize}
\end{minipage}}\\
\medskip

\noindent
These different cases are illustrated in \textbf{Figure\,\ref{fig4}(b)}. Note that we have the correspondence (\ref{eq:Verdet_constant}) for the Verdet constant. Thus, showing that the effective Verdet constant cannot be enhanced is equivalent to all points $(\sigma_*,\,|S_*|)$ lying below the curve $\sqrt{\sigma_{\perp}^*}\max_i|S_i|/\sqrt{\sigma_i}$ in the $(\sigma_{\perp}^*,\,|S_{\parallel}^*|)$-plane, cf. \textbf{Figure\,\ref{fig4}(a)}. Note that, as the result holds for the Verdet constant of uniaxial metamaterials, it also holds for isotropic metamaterials.\\
\medskip

\noindent
The third conclusion again follows from considering the bounds segment-wise. For the straight-line bounds [case (i)], we obtain that the maximum is attained by the right endpoint. For the parabolic-arc bounds, we obtain that, in case (ii)(a), the left endpoint attains the maximum, whereas in case (ii)(b) it is the right endpoint that attains the maximum.\\
\medskip

\noindent
While the result for the Verdet constant is not as general as the corresponding statements for the Hall effect, it should be pointed out that there are several ways of obtaining improvements: First, using translations, one can extend the statement to many other cases. Second, as we will show below, in the two-phase case, there is no enhancement irrespective of the properties of the constituent materials.\\
\medskip

\noindent
Potentially, one could go yet farther: Indeed we hypothesize that, irrespective of the number of phases, the Verdet constant cannot be enhanced in the quasistatic non-resonant regime. One potential avenue to show this is to use the methods that we use for the two-phase case below. While these methods generalize to the multiphase case, the analysis becomes significantly more involved.

\section{Improvements in the two-phase case}\label{sec5}

\noindent
Improvements over the results presented above can be obtained with relative ease if the metamaterial is made from two phases only. In the following, we will present corresponding bounds and discuss their attainability. In contrast to the multiphase case, we will also present bounds that take the volume-fraction of the phases into account. These bounds are subsequently used to find their volume-fraction independent counterparts. Finally, the improved lower bound will allow us to draw stronger conclusions. Specifically, we will show that the effective Verdet constant cannot be enhanced in two-phase metamaterials and will identify the material contrast that is necessary to obtain sign-inversions.

\subsection{The two-phase bounds and their attainability}

\noindent
In this section, we consider uniaxial metamaterials made from two isotropic phases. The phases have zmf\,conductivities $\sigma_1$ and $\sigma_2$ and $S$-coefficients $S_1$ and $S_2$. For simplicity, we will set $S_2=0$ (corresponding results for $S_2\neq0$ follow readily through a suitable translation). In our usage of the terms ``lower'' and ``upper'' bound, we assume that $S_1>0$, which corresponds to the condition $S_1>S_2$ if we allow for arbitrary values of $S_2$. Otherwise, the bounds flip roles. Instead of working with the effective zmf\,conductivity, $\sigma_{\perp}^*$, and axial effective $S$-coefficient, $S_{\parallel}^*$, we will express the bounds and effective properties in terms of the components of the corresponding $Y$-tensor, which is given by
\begin{align}
\widetilde{\bm{Y}} =\begin{pmatrix} y_{\sigma\perp} & -y_{S} & 0 \\ 
	y_{S} & y_{\sigma\perp} & 0 \\
	0 & 0 & y_{\sigma\parallel}
\end{pmatrix}
\quad \text{with} \quad
{\small
	\begin{aligned}
		y_{\sigma\perp}+iy_{S} &=-f_2(\sigma_{1}+iS_1B)-f_1\sigma_{2}+\frac{f_1f_2(\sigma_{1}+iS_1B-\sigma_{2})^2}{f_1(\sigma_{1}+iS_1B)+f_2\sigma_{2}-(\sigma_{\perp}^*+iS_{\parallel}^*B)}, \\
		y_{\sigma\parallel} &= -f_2\sigma_{1}-f_1\sigma_{2}+\frac{f_1f_2(\sigma_{1}-\sigma_{2})^2}{f_1\sigma_{1}+f_2\sigma_{2}-\sigma_{\parallel}^*}.
	\end{aligned}
}
\end{align}
A key advantage of using the $Y$-tensor is that in terms of $\widetilde{\bm{Y}}$ the expressions for our bounds are volume-fraction independent and that the effective properties of many microstructures become significantly simpler. For further details, see the derivation of the bounds below.\\
\medskip

\begin{figure}[h!]
	\centering
	\makebox[\textwidth]{\includegraphics{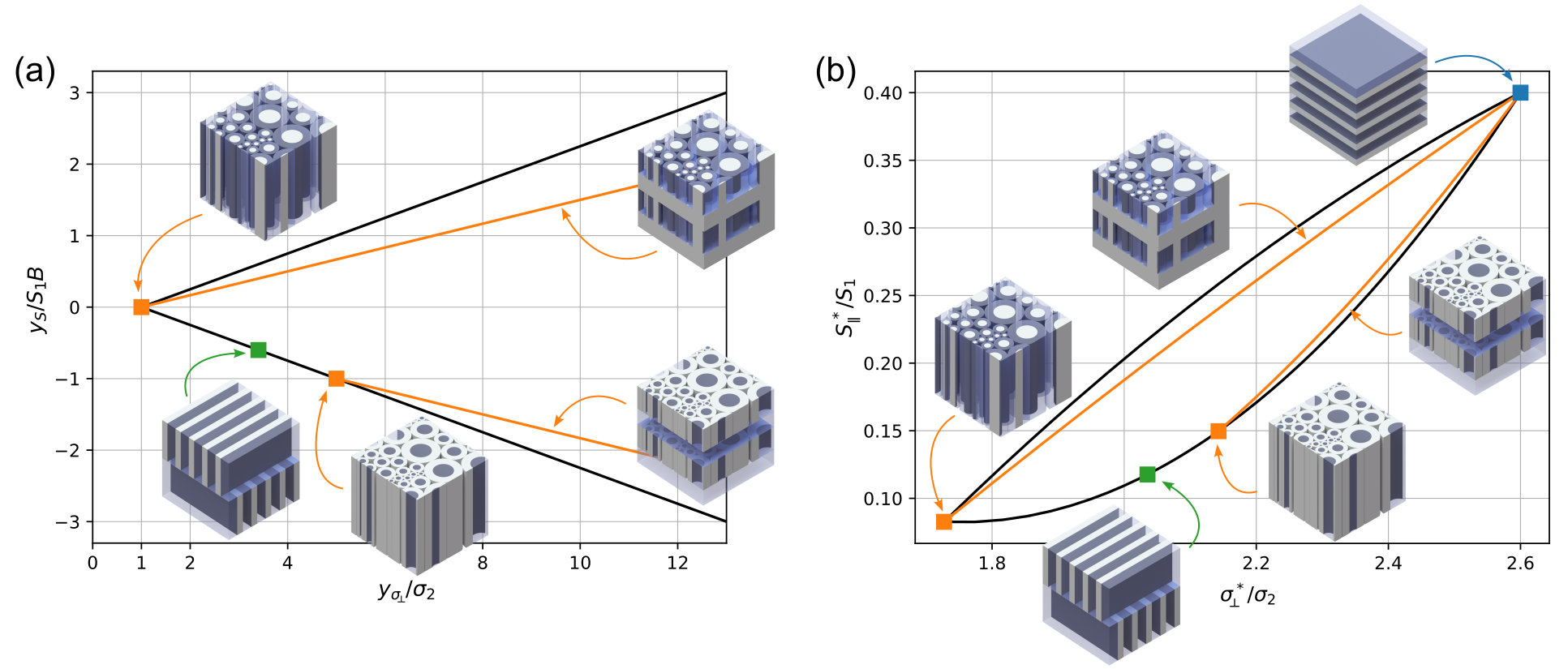}}
	\caption{Volume-fraction dependent bounds relating the axial $S$-coefficient to the transversal conductivity of uniaxial two-phase metamaterials. Via the polycrystal argument discussed in Section\,\ref{secproperties}, the bounds apply more generally to fully anisotropic metamaterials. Panels (a) and (b) show the bounds in the $y$-plane and the effective parameter plane, respectively. The colored squares correspond to different microstructures attaining the bounds, which are illustrated via the corresponding insets. Specifically, the orange squares correspond to assemblages of coated cylinders while the blue and the green square correspond to a rank-$1$ laminate and a specific polycrystal formed from a rank-$1$ laminate, respectively. The orange lines correspond to coated cylinders assemblage that are layered with the respective core phase perpendicular to the cylinder axes. The mathematical analysis assumes infinitely separated length-scales for the hierarchical microstructures (the polycrystal and the layered coated cylinder assemblages), contrary to the depiction. Furthermore, the coated cylinders are assumed to fill the entire space. Parameters are $\sigma_1/\sigma_2=5$, $S_2=0$, and $f_1=0.4$.}
	\label{fig5}
\end{figure}

\noindent
In terms of these $y$-parameters, the admissible range of effective parameters corresponds to the wedge defined by
\begin{equation}
	|y_S| \leq \frac{y_{\sigma\perp}-\sigma_2}{\sigma_1-\sigma_2}|S_1B|,
	\label{eq:boundtwophase}
\end{equation}
i.e., the wedge is bounded by two straight lines passing through the points $(\sigma_2,~0)$ and $(\sigma_1,~S_1B)$ or\\ $(\sigma_1,~-S_1B)$. In the $(\sigma_{\perp}^*,\,S_{\parallel}^*)$-plane, these bounds become arcs that form a convex lens-shaped region, which morphs and moves as the volume fraction is varied. An illustration of the bounds in both the $y$-plane and the effective parameter plane is shown in \textbf{Figure\,\ref{fig5}} -- together with the microstructures attaining (or approaching) the bounds discussed below. Analytic expressions for the bounds in the effective parameter plane are given in the Supporting Information.\\
\medskip

\noindent
Four points on the bounds are attained by the following microstructures: The first two points (the orange squares in Figure\,\ref{fig5}) correspond to coated cylinder assemblages, which are formed in the way this term suggests: Consider an (infinitely extended) cylinder made from one of the phases (the ``core phase''). Next, this cylinder is coated with the other phase (the ``coating phase''), i.e., a cylindrical shell made from the other phase is added. Finally, scaled copies of this coated cylinder are assembled with aligned axes in such a way that the entire space is filled. For phase $1$ as the core phase, one obtains the following $y$-parameters
\begin{equation}
	y_{\sigma\perp}^{\text{CC1}} = \sigma_2,~y_S^{\text{CC1}}=0.
\end{equation}
Thus, this first coated cylinder assemblage corresponds to the vertex of the wedge in the $y$-plane. In the effective parameter plane, it attains the lower intersection of the two arcs.
Using phase $2$ as the core phase, one instead obtains
\begin{equation}
	y_{\sigma\perp}^{\text{CC2}} = \sigma_1,~y_S^{\text{CC2}}=-S_1B.
\end{equation}
Thus, the second coated cylinder assemblage attains a point on the lower bound.\\
\medskip

\noindent
The third microstructure is the uniaxial polycrystal discussed in the context of the multiphase bounds above. The $y$-parameters of this polycrystal are
\begin{equation}
	y^{\text{L}} =  f_1y^{\text{CC1}}+f_2y^{\text{CC2}},
\end{equation}
which holds for all components. As the volume fraction is varied, this point (the green square in Figure\,\ref{fig5}) moves along the lower bound between the points corresponding to the coated cylinder assemblages. Thus, this section of the bounds is the tightest possible bound that is volume-fraction independent in the $y$-plane.\\
\medskip

\noindent
The fourth point (blue square in Figure\,\ref{fig5}(b)) corresponds to a rank-$1$ laminate. Its effective properties are given by (\ref{eq:effproplam}). Note that this point lies at $y_{\sigma\perp}\rightarrow\infty$ in the $y$-plane.\\
\medskip

\noindent
Additionally, it is possible to interpolate between the two coated cylinder assemblages and the rank-$1$ laminate in a natural way by laminating the coated cylinder assemblages with their respective core phase with the layers being perpendicular to the cylinder axes (generating the orange lines and curves in Figure\,\ref{fig5}). With phase~$1$ as the core phase, the effective $y$-parameters of this structure are
\begin{equation}
	y_{\sigma\perp} = \frac{\left(f_1-f^{\text{CC}}_1\right)\sigma_1+f_1\sigma_2}{f^{\text{CC}}_1},~y_S=\frac{f_1-f^{\text{CC}}_1}{f^{\text{CC}}_1}S_1B,
	\label{eq:CClayered}
\end{equation}
where $f^{\text{CC}}_1$ is the volume fraction of phase $1$ in the coated cylinder assemblage. For $f^{\text{CC}}_1=0$, one recovers the rank\nobreakdash-$1$ laminate, whereas for $f^{\text{CC}}_1=f_1$, the structure reduces to the coated cylinder assemblage. The corresponding phase-interchanged structure, in which phase~$2$ is the core phase, has effective $y$-parameters
\begin{equation}
	y_{\sigma\perp} = \frac{f_2\sigma_1+\left(f_2-f^{\text{CC}}_2\right)\sigma_2}{f^{\text{CC}}_2},~y_S=-\frac{f_2}{f^{\text{CC}}_2}S_1B,
	\label{eq:CClayeredinv}
\end{equation}
where $f^{\text{CC}}_2$ is the volume fraction of phase $2$ in the coated cylinder assemblage. While these structures generally do not attain the bounds, they come arbitrarily close to doing so if the conductivity ratio is sufficiently large, as can be seen by considering the limit $\sigma_2/\sigma_1\rightarrow 0$ in Equations\,(\ref{eq:CClayered}) and (\ref{eq:CClayeredinv}). As illustrated in Figure\,\ref{fig5}, for a conductivity ratio of $\sigma_1/\sigma_2=5$ the microstructures are already reasonably close to attaining the bounds. Note that we have not succeeded in identifying a hierarchical laminate microstructure that improves on the layered coated cylinder assemblages. We thus speculate that they are in fact optimal.\\
\medskip

\begin{figure}[h!]
	\centering
	\includegraphics{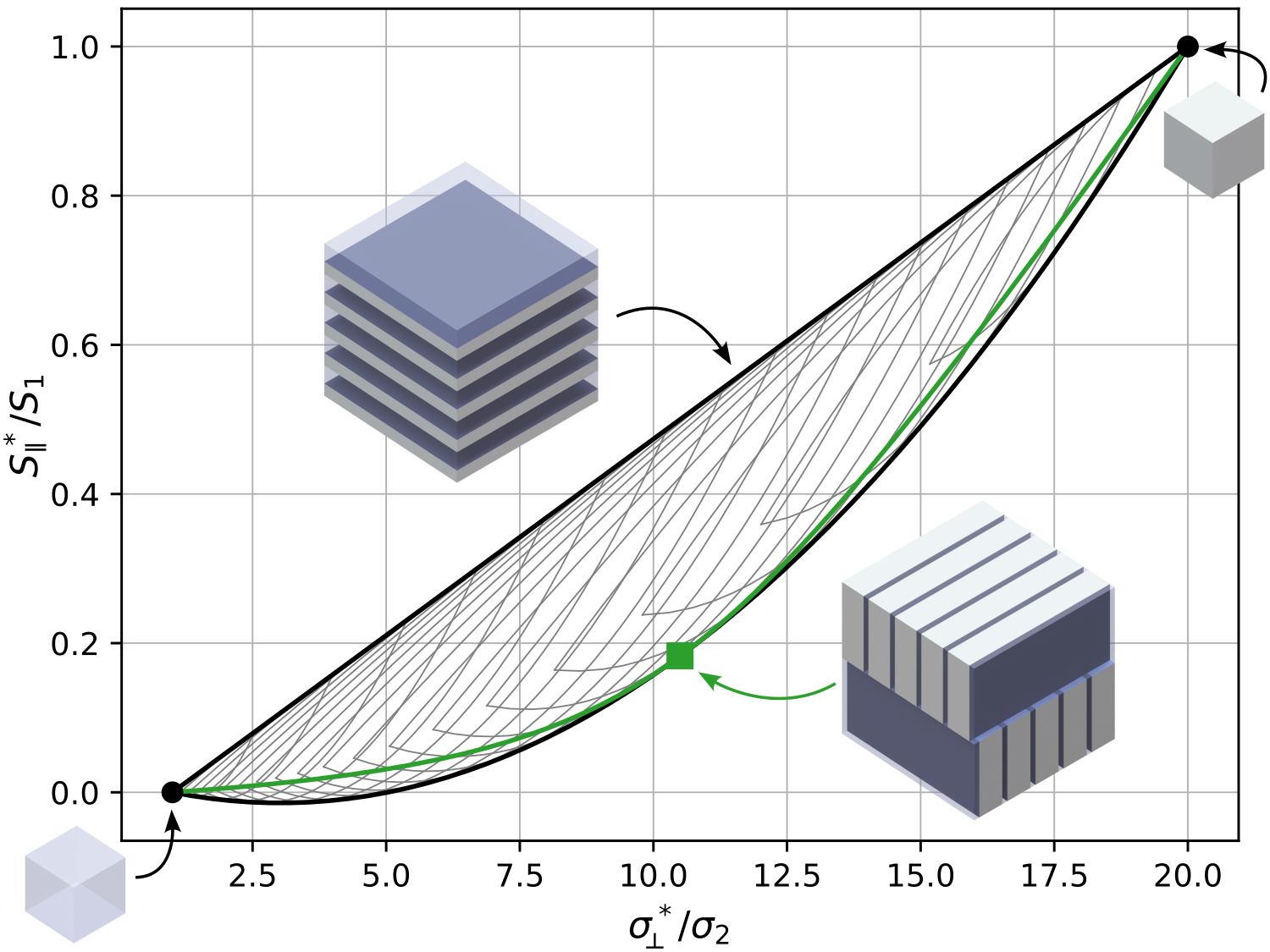}
	\caption{Volume-fraction independent two-phase bounds relating the axial $S$-coefficient to the transversal conductivity of a uniaxial two-phase metamaterial. The upper and lower bound are a straight line and a parabola, respectively. The vertices at the intersections correspond to the pure phases. The thin gray lines correspond to the volume-fraction dependent bounds for a number of selected volume fractions between zero and one. The upper bound is in its entirety attained by a rank-$1$ laminate. The green line corresponds to the polycrystal formed from a rank-$1$ as described in the main text. At a specific volume fraction, this polycrystal attains the lower bound at the point marked by the green square. At other volume fractions, it comes reasonably close to doing so. Unlike shown, the mathematical analysis assumes that the length-scales on which the polycrystal is formed is much larger than the unit cell of the underlying rank-$1$ laminate. Parameters are $\sigma_1/\sigma_2=20$ and $S_2=0$.}
	\label{fig6}
\end{figure}

\noindent
From the volume-fraction dependent bounds, one can readily obtain their volume-fraction independent counterparts. To do so, for each value of the effective conductivity, and for the upper and lower bounds separately, one can find the volume-fraction dependent bound imposing the weakest constraint on the effective axial $S$-tensor component. The resulting volume-fraction independent bounds are illustrated in \textbf{Figure\,\ref{fig6}}. Note that the volume-fraction independent bounds naturally inherit the property to apply to structures that do not satisfy any symmetry requirements upon making the replacements (\ref{eq:poStens}).\\
\medskip

\noindent
For the upper bound, we recover the two-phase version of the bound (\ref{eq:bound_sl}), i.e., the upper bound is the straight line passing through the points $(\sigma_2,~0)$ and $(\sigma_1,~S_1B)$ corresponding to the two pure phases. As discussed in detail above, the entirety of this bound is attained by a rank-$1$ laminate.

\medskip 

\noindent
A substantial improvement over the multiphase bounds is obtained for the volume-fraction independent lower bound. This improved bound is given by the parabola that, in addition to passing through the two points corresponding to the pure phases, has a minimum at 
\begin{equation}
	\sigma_{\perp}^*=\frac{1}{4}\left(\sigma_1+\sigma_2-2\sqrt{\sigma_1\sigma_2}\right).
	\label{eq:lowerboundmin}
\end{equation}
In Section\,\ref{sec_twophaseimpl} below, we will use this bound to show that there are no enhancements of the effective (axial) Verdet constant in the two-phase case. Furthermore, we will use it to draw conclusions on the conductivity contrast required for sign-inversions.\\
\medskip

\noindent
One point (corresponding to the green square in Figure\,\ref{fig5}) on the lower bound is attained by the polycrystal formed from a rank-$1$ laminate according to the procedure described in detail above. The volume fraction for which this polycrystal attains the bound is $f_1 = \sigma_1/(\sigma_1+\sqrt{\sigma_1\sigma_2})$. More generally, the curve traced out by the polycrystal in the effective parameter plane as the volume fraction is varied comes reasonably close to the lower bound. This holds in particular for relatively small values of the conductivity ratio.\\
\medskip

\noindent
Note that this polycrystal fundamentally cannot exhibit a sign-inversion (as it is an orthogonal laminate \cite{Briane:2004:WKL,Kern:2023:SRE}). Several microstructures exhibiting a sign-inversion of (at least) one diagonal component of the effective $S$-tensor -- and, thus, approaching the bounds more closely than the polycrystal in this regime -- are known \cite{Briane:2009:HTD,Kern:2018:THE,Kern:2023:SRE}. The microstructure showing the strongest sign-inversion identified hitherto is (to the best of our knowledge) the rank-$3$ laminate introduced in \cite{Kern:2023:SRE}. We will further elaborate on the particularly interesting sign-inverted regime in section below.

\subsection{Derivation of the two-phase bounds}

\noindent
To derive the two-phase bounds, we will again use the monotonicity argument discussed in Section\,\ref{sec3}. The difference to the multiphase case and key for the improvement is, instead of working with the effective tensor $\bm{\sigma}^{*+}$, to consider a sequence of effective tensors relating the fluctuations of successively higher order in the electric field and the electric current density to one another. In our case, two steps in this sequence will suffice. This method is known as the field-equation recursion method. For details, we refer to Refs.\,\cite{Milton:1987:MCEa,Milton:1987:MCEb,Milton:1991:FER,Clark:1997:CFR,Milton:1985:TCC} and Chapter\,29 in \cite{Milton:2002:TOC}.\\
\medskip

\noindent
We start from the effective conductivity tensor, $\bm{\sigma}^{*+}$, which relates the averages of the fields, $\langle\bm{j}\rangle=\bm{\sigma}^{*+}\langle\bm{e}\rangle$. In the first step, we subtract the averages from the electric field and the electric current density. We obtain a homogenization problem that shares key properties with the original one but in which the averages of the fields vanish. Associated with this homogenization problem is a sort of effective tensor $\bm{Y}^+$ which, in each of the phases, relates the fluctuations of the fields to one another,
\begin{equation}
	\left\langle\chi_i(\bm{j}-\langle\bm{j}\rangle)\right\rangle=-\bm{Y}^+\left\langle\chi_i(\bm{e}-\langle\bm{e}\rangle)\right\rangle \text{for }i=1,2.
\end{equation}
In terms of the conductivity tensors of the two constituent materials, $\bm{\sigma}_1^+$ and $\bm{\sigma}_2^+$, and the effective conductivity tensor of the metamaterial, $\bm{\sigma}^{*+}$, the $Y^+$-tensor is given by
\begin{equation}
	\bm{Y}^+=-f_2\bm{\sigma}_1^+-f_1\bm{\sigma}_2^++f_1f_2(\bm{\sigma}_1^+-\bm{\sigma}_2^+)(f_1\bm{\sigma}_1^++f_2\bm{\sigma}_2^+-\bm{\sigma}^{*+})^{-1}(\bm{\sigma}_1^+-\bm{\sigma}_2^+).
\end{equation}
The advantage of introducing $\bm{Y}^+$ is that elementary bounds on this tensor, which can be derived based on the properties that $\bm{Y}^+$ inherits (specifically homogeneity, Herglotz, and normalization properties), get mapped to more complicated bounds on $\bm{\sigma}^{*+}$. In particular, bounds on $\bm{Y}^+$ that are volume fraction independent get mapped to bounds on $\bm{\sigma}^{*+}$ that are volume fraction dependent.\\
\medskip

\noindent
Going one step further, one can eliminate fields that are constant in each phase. The associated effective tensor $\bm{\sigma}^{*(1)+}$ is related to $\bm{Y}^+$ via
\begin{equation}
	\bm{\sigma}^{*(1)+}=\bm{N}^{-1/2}\bm{Y}^{+}\bm{N}^{-1/2},
	\label{eq:firstordereff}
\end{equation}
wherein the normalization $\bm{N}$, which is the value of $\bm{Y}^{+}$ when $\sigma_1^+=\sigma_2^+=1$, depends on the second derivative of the effective conductivity. Furthermore, it is well-known that information about the symmetry of a metamaterial is reflected in this derivative \cite{Bergman:1978:DCC}. Thus, information about the symmetry of a metamaterial can be incorporated into the bounds through the constraints that they pose on the normalization in (\ref{eq:firstordereff}).\\
\medskip

\noindent
Specifically for two-constituent uniaxial metamaterials, the effective transversal conductivity satisfies \cite{Milton:1981:BCP}
\begin{equation}
	\frac{\partial^2\sigma_{\perp}^{*+}(\sigma_1,1)}{\partial \sigma_1^2}\geq-f_1f_2,
\end{equation}
where $f_1$ and $f_2$ are the volume-fractions of the two constituent materials. Using this constraint on the derivative, we obtain $n_{\perp}\geq 1$ wherein $n_{\perp}$ is the transversal normalization factor, i.e., $n_{\perp}\sigma_{\perp}^{*(1)+}=y_{\perp}^+$.\\
\medskip

\noindent
Applying our bounds (\ref{eq:bound_sl}) to $\sigma_{\perp}^{*(1)+}$, which we can do as the associated homogenization problem shares the necessary properties with the original one, we obtain
\begin{equation}
	n_{\perp}^{-1}(y_{\perp}^+-y_{\perp})=\sigma_{\perp}^{*(1)+}-\sigma_{\perp}^{*(1)}\leq |S_1|\frac{\sigma_{\perp}^{*(1)}-\sigma_2}{\sigma_1-\sigma_2} = |S_1|\frac{n_{\perp}^{-1}y_{\perp}-\sigma_2}{\sigma_1-\sigma_2},
\end{equation}
where $\sigma_{\perp}^{*(1)}$ and $y_{\perp}$ are the transversal components of the corresponding zero magnetic field tensors. Using the monotonicity result $|y_S|\leq y_{\perp}^+-y_{\perp}$ and the constraint on the normalization factor, we thus obtain the bound (\ref{eq:boundtwophase}), which concludes the derivation.

\subsection{Key implications -- two-phase case}\label{sec_twophaseimpl}

\noindent
Using the lower volume-fraction independent bound, we can now draw improved conclusions. The first conclusion concerns the effectiveVerdet constant: In the multiphase case, we found that the Verdet constant cannot be enhanced if the properties of the pure phases satisfy certain additional restrictions [cases (i) and (ii)(a)(b) above]. In contrast, for two-phase metamaterials, these restrictions can be lifted, i.e., the implication $|\mathcal{V}_{\parallel}^*|\leq \max \left\{|\mathcal{V}_1|,|\mathcal{V}_2|\right\}$ holds irrespective of the properties of the pure phases. We emphasize, however, that the assumptions underlying the derivation of the bounds, in particular, the absence of resonances, remain in place. This conclusion holds for arbitrary values of $\gamma_2$, not just in the case $\gamma_2=0$. Note that, to arrive at the desired conclusion, we found it helpful to use the estimate $|S_{\parallel}^*|/\sqrt{\sigma_{\perp}^*} \leq |S_{\parallel}^*|(\sqrt{\sigma_1}+\sqrt{\sigma_2})/(\sqrt{\sigma_1\sigma_2}+\sigma_{\perp}^*)$, which follows from $\sigma_{\perp}^*$ lying between $\sigma_1$ and $\sigma_2$.

\medskip

\noindent
The second conclusion concerns the particularly interesting sign-inverted regime. Considering the minimum of the parabola, we obtain that, in the case $S_2=0$, the conductivity ratio has to satisfy $\sigma_1 > 9\sigma_2$ for sign-inversions of $S_{\perp}^*$ not to be ruled out. For $S_2 \neq 0$, this is the condition for $S_{\perp}^*<S_2$ (if $S_1>S_2>0$) or $S_{\perp}^*>S_2$ (if $S_1<S_2<0$). Via the replacement (\ref{eq:poStens}) following from the polycrystal argument, these conditions apply more generally to diagonal components of the effective $S$-tensor of an anisotropic metamaterial. This is in line with the general observation in the theory of metamaterials that unusual effective properties require a sufficiently large contrast in the properties of the phases.

\medskip

\noindent
Note that the same conditions apply to the effective axial Hall tensor component of uniaxial metamaterials, as this quantity is sign-inverted if and only if the axial $S$-tensor component is sign-inverted. Via the replacement (\ref{eq:faeq}), these conditions extend to diagonal components of the effective Hall-tensor in the general anisotropic case.

\medskip

\begin{figure}[h!]
	\centering
	\includegraphics{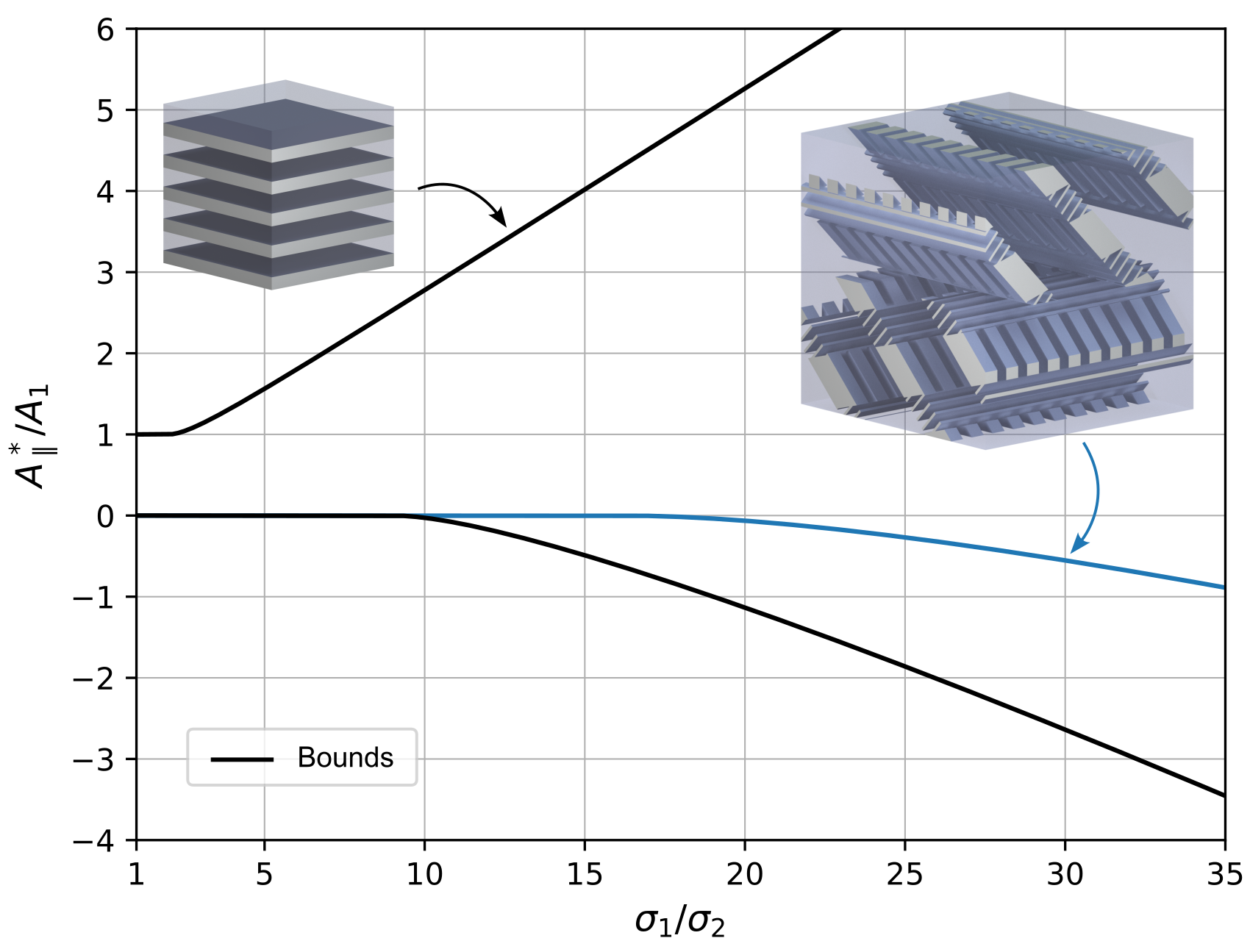}
	\caption{Bounds on the effective axial Hall tensor component (black lines) of a two-phase uniaxial metamaterial as a function of the conductivity contrast, $\sigma_1/\sigma_2$. The upper bound is attained by a rank-$1$ laminate (which reduces to pure phase $1$ for $\sigma_1/\sigma_2\leq 2$). The blue line corresponds to an optimized uniaxial polycrystal formed from the anisotropic rank-$3$ laminate introduced in \cite{Kern:2023:SRE}. The length-scales of subsequent lamination steps are assumed to be infinitely separated, contrary to the depiction.}
	\label{fig7}
\end{figure}

\noindent
More generally, we can use the two-phase bounds to establish limits on the range of the diagonal components of the effective Hall tensor as a function of the conductivity contrast of the phases. These bounds are illustrated in \textbf{Figure\,\ref{fig7}}. Explicit expressions are provided in the Supporting Information. The general pattern is the same for the upper and lower bound. At low conductivity contrasts, the pure phases are optimal. Above certain thresholds, the bounds allow for enhancements, $A_{\parallel}^*/A_1>1$, and sign-inversions, $A_{\parallel}^*\cdot A_1<0$. Notably, the threshold for a sign-inversion lies much higher than the respective threshold for an enhancement (which is exactly $\sigma_1 > 2 \sigma_2$).

\medskip

\noindent
As already discussed in the context of the multiphase case, the upper bound is attained by a rank-$1$ laminate. Note that the layers of this laminate are perpendicular to the axial direction (or more generally perpendicular to the $x_i$-direction for a bound on the component $A_{ii}^*$).

\medskip

\noindent
Whether the lower bound is attained remains an open problem. The microstructure with the strongest sign-inversion identified so far is, to the best of our knowledge, a polycrystal formed from the rank-$3$ laminate introduced in \cite{Kern:2023:SRE}. The polycrystal is formed by laminating this microstructure with a rotated copy of itself ($90^{\circ}$ rotation about the axial direction and layers perpendicular to the same axis). The microstructure achieves a sign-inversion through a reversal of the microscopic direction of current flow. This local current-inversion is obtained through several lamination steps at oblique angles, which cause a successive rotation of the current density. Note that using oblique lamination angles is essential: orthogonal laminates can fundamentally not exhibit a sign-inversion \cite{Kern:2023:SRE,Briane:2004:WKL}. Furthermore, the current reversal itself is not sufficient. Additionally, one has to translate the inversion of a local Hall voltage into a corresponding macroscopic effect. In Figure\,\ref{fig7}, we show the results of a minimization of the axial component of the effective Hall tensor over the geometry parameters of the laminate (see Section\,IV of Ref.\,\cite{Kern:2023:SRE} for details). While the structure exhibits a sign-inversion for conductivity contrasts as low as $\sigma_1/\sigma_2=16.2$, this value is still significantly larger than the smallest possible threshold $\sigma_1 > 9\sigma_2$. Furthermore, there is still a significant gap between the achieved magnitude of the inversion and the bound. Whether this gap is due to limitations of the microstructure, the bound, or both, remains to be investigated further.

\section{Conclusion}\label{sec6}

\noindent
In this paper, we have studied the limits to nonreciprocal effects in a wide range of metamaterials. By establishing tight bounds and identifying microstructures that attain or closely approach them, we have gained a substantially refined picture of the effective nonreciprocal properties achievable through the powerful ``metamaterial'' paradigm of judiciously tailoring unit cell geometry.

\medskip

\noindent
Our analysis rests on three key assumptions: (i) the antisymmetric contribution to the material tensors is small, (ii) the governing equations are equivalent to the conductivity problem (restricting wave phenomena to the quasistatic regime), and (iii) the unperturbed (``zero magnetic-field'') tensor is self-adjoint and positive definite.

\medskip

\noindent
Under these assumptions, we have derived a set of multiphase bounds that are at least partly optimal. These bounds allowed us to conclude that there are no enhancements of the effective Hall mobility and (under additional assumptions on the properties of the phases) the effective Verdet constant, and that the largest value of diagonal Hall tensor components is attained either by one of the pure phases or by a rank-$1$ laminate formed from two of the pure phases.

\medskip

\noindent
In the two-phase case, we have obtained both volume-fraction dependent and independent bounds that are at least partly optimal. The latter bounds allowed us to conclude that there are no enhancements of the effective Verdet constant in two-phase metamaterials regardless of the properties of the phases. Furthermore, they enabled us to identify a conductivity contrast that is necessary for sign-inversions.

\medskip

\noindent
We expect that the presented approach will continue to prove useful for understanding the effective properties of nonreciprocal metamaterials -- particularly as relying on the monotonicity of the effective tensor enables the use of many powerful techniques originally derived for conventional (i.e., reciprocal) metamaterials. A particularly interesting avenue is to extend the analysis of the Faraday effect to allow for absorption and plasmonic resonances, e.g., by applying the Cherkaev-Gibiansky transformation \cite{Cherkaev:1994:VPC}. Another central question is whether the bounds can be extended beyond the quasistatic regime, potentially by building on the profound recent progress in this area \cite{Miller:2015:FLO}.

\newpage
\noindent
\textbf{Acknowledgements} \par 
\noindent
The authors are grateful to the National Science Foundation for support through the Research Grant No. DMS-2107926. C. Kern acknowledges financial support from the Villum Foundation through the Villum Investigator Project Amstrad (VIL54487).

\medskip

\ifx \bblindex \undefined \def \bblindex #1{} \fi\ifx \bbljournal \undefined
\def \bbljournal #1{{\em #1}\index{#1@{\em #1}}} \fi\ifx \bblnumber
\undefined \def \bblnumber #1{{\bf #1}} \fi\ifx \bblvolume \undefined \def
\bblvolume #1{{\bf #1}} \fi\ifx \noopsort \undefined \def \noopsort #1{}
\fi\ifx \bblindex \undefined \def \bblindex #1{} \fi\ifx \bbljournal
\undefined \def \bbljournal #1{{\em #1}\index{#1@{\em #1}}} \fi\ifx
\bblnumber \undefined \def \bblnumber #1{{\bf #1}} \fi\ifx \bblvolume
\undefined \def \bblvolume #1{{\bf #1}} \fi\ifx \noopsort \undefined \def
\noopsort #1{} \fi

\clearpage

\setcounter{page}{1}
\renewcommand{\theequation}{S\arabic{equation}}
\renewcommand{\thesection}{S\Roman{section}}
\renewcommand{\thesubsection}{\Alph{subsection}}

\setcounter{section}{0}
\setcounter{subsection}{0}

\begin{center}
	
	{\large\textbf{Supporting Information:} Limits to the Hall effect and \\other nonreciprocal effects in three-dimensional metamaterials}
	
	\vspace{1.5em}
	
	{\normalsize
		Christian Kern$^{1}$ and Graeme W. Milton$^{2}$\\
	}
	\vspace{1em}
	
	{\small\it
		$^{1}$Department of Civil and Mechanical Engineering,\\ Technical University of Denmark, 2800 Kgs. Lyngby, Denmark\\
		$^{2}$Department of Mathematics, University of Utah, Salt Lake City, UT 84112, USA
	}
	
\end{center}

\begin{center}
Email Addresses: physics@chrkern.de, graeme.milton@utah.edu
\end{center}

\vspace{2em}

\makeatletter
\newcommand{\sicontents}{
	\@starttoc{sitoc}
}
\newcommand{\siaddcontentsline}[3]{
	\addcontentsline{sitoc}{#1}{#2}
}
\makeatother
\let\oldsection\section
\let\oldsubsection\subsection
\renewcommand{\section}[1]{
	\oldsection{#1}
	\siaddcontentsline{section}{\makebox[2.5em][l]{\thesection.}#1}{}
}
\renewcommand{\subsection}[1]{
	\oldsubsection{#1}
	\siaddcontentsline{subsection}{\hspace{2.0em}\makebox[2.5em][l]{\thesubsection.}#1}{}
}

\begin{center}
\noindent\textbf{CONTENTS}
\end{center}
\sicontents

\newpage

\section{Extending the bounds to fully anisotropic materials}

\noindent
In this section, we continue the discussion from Section 2.1 of the main text and show how the bounds on uniaxial metamaterials yield important implications for (generally) fully anisotropic metamaterials. The idea is to form uniaxial microstructures (involving several lamination steps and the formation of a polycrystal) from the fully anisotropic metamaterials. Applying bounds to these uniaxial microstructures then implies bounds on the effective properties of fully anisotropic metamaterials.\\

\noindent
Specifically, on the basis of a first such microstructure, we demonstrate in Subsection \ref{Halltenscomp} that the uniaxial bounds apply to fully anisotropic metamaterials upon making the replacements in Equation\,(17) of the main text, which we repeat here
\begin{equation*}
	\sigma_{\perp}^* \rightarrow \sqrt{\text{det}\left(\bm{\sigma}^{*}\right)/\sigma_{33}^{*}} \text{ and } A_{\parallel}^{*} \rightarrow A_{33}^{*}.
\end{equation*}
On the basis of a second such microstructure, we establish in Subsection \ref{Stenscomp} that the uniaxial bounds also apply to fully anisotropic metamaterials upon making the alternative replacements in Equation\,(18) of the main text,
\begin{equation*}
	\sigma_{\perp}^{*} \rightarrow \sqrt{\sigma_{11}^{*}\sigma_{22}^{*}-\left.\sigma_{12}^{*}\right.^2}=\sqrt{\text{Cof}\left(\bm{\sigma}^{*}\right)_{33}} \text{ and } S_{\parallel}^{*} \rightarrow S_{33}^{*}.
\end{equation*}

\subsection{Hall tensor components}\label{Halltenscomp}

\noindent
Consider a fully anisotropic metamaterial with effective zero magnetic-field (zmf) conductivity tensor $\bm{\sigma}^*$ and effective $S$-tensor $\bm{S}^*$. In addition to this metamaterial, we consider a copy that is mirrored at the $x_1x_2$-plane. This mirror copy has effective properties
\begin{align}
	\bm{\sigma}^{\text{(*M)}}=\begin{pmatrix} \sigma_{11}^* && \sigma_{12}^* && -\sigma_{13}^* \\
		\sigma_{12}^* && \sigma_{22}^* && -\sigma_{23}^* \\
		-\sigma_{13}^* && -\sigma_{23}^* && \sigma_{33}^* \end{pmatrix} \text{ and }		
	\bm{S}^{\text{(*M)}}=\begin{pmatrix} S_{11}^* && S_{12}^* && -S_{13}^* \\
		S_{21}^* && S_{22}^* && -S_{23}^* \\
		-S_{31}^* && -S_{32}^* && S_{33}^* \end{pmatrix}.
\end{align}	
Next, we laminate the original metamaterial and its mirrored copy in equal proportions with layers perpendicular to the $x_3$-direction. Using the usual lamination equations and the perturbation expression (12), we obtain the following expression for the effective conductivity tensor and the effective $S$-tensor,
\begin{align}
	\bm{\sigma}^{\text{(*2)}}=\begin{pmatrix} \sigma_{11}^*-\frac{\left(\sigma_{13}^*\right)^2}{\sigma_{33}^*} && \sigma_{12}^*-\frac{\sigma_{13}^*\sigma_{23}^*}{\sigma_{33}^*} && 0 \\
		\sigma_{12}^*-\frac{\sigma_{13}^*\sigma_{23}^*}{\sigma_{33}^*} && \sigma_{22}^*-\frac{\left(\sigma_{23}^*\right)^2}{\sigma_{33}^*} && 0 \\	0 && 0 && \sigma_{33}^* \end{pmatrix} \text{ and }
	\bm{S}^{\text{(*2)}}=\begin{pmatrix} S_{11}^* && S_{12}^* && 0 \\
		S_{21}^* && S_{22}^* && 0 \\
		0 && 0 && \frac{\sigma_{13}^*}{\sigma_{33}^*}S_{13}^*+\frac{\sigma_{23}^*}{\sigma_{33}^*}S_{23}^*+S_{33}^* \end{pmatrix}.
\end{align}	
We now perform a second lamination. This time we laminate in equal proportions with layers perpendicular to the $x_1$-direction. The laminate is formed from the laminate from the previous step and a copy of it that is mirrored at the $x_1x_3$-plane. The effective properties of the resulting metamaterial are
\begin{align}
	\bm{\sigma}^{\text{(*3)}} &= \text{diag}\left( \sigma_{11}^*-\frac{\left(\sigma_{13}^*\right)^2}{\sigma_{33}^*}, \sigma_{22}^*-\frac{\left(\sigma_{23}^*\right)^2}{\sigma_{33}^*}-\left(\sigma_{12}^*-\frac{\sigma_{13}^*\sigma_{23}^*}{\sigma_{33}^*}\right)^2\left(\sigma_{11}^*-\frac{\left(\sigma_{13}^*\right)^2}{\sigma_{33}^*}\right)^{-1}, \sigma_{33}^*\right)\\ \text{ and }
	\bm{S}^{\text{(*3)}}&=\text{diag}\left( S_{11}^*, S_{22}^*+S_{12}^*\left(\sigma_{12}^*-\frac{\sigma_{13}^*\sigma_{23}^*}{\sigma_{33}^*}\right), \frac{\sigma_{13}^*}{\sigma_{33}^*}S_{13}^*+\frac{\sigma_{23}^*}{\sigma_{33}^*}S_{23}^*+S_{33}^* \right).
\end{align}	
In the last step, we form a uniaxial polycrystal from this metamaterial. Specifically, we construct an extruded (columnar) version of a checkerboard. The squares of one of the two colors (e.g., the white squares) are made from the metamaterial formed in the previous step. In the squares of the other color (e.g., the black squares), the metamaterial is rotated by $90^{\circ}$ about the $x_3$-axis. This polycrystal has effective properties 
\begin{align}
	\sigma_{\perp}^{\text{(*P)}} = \sqrt{\text{det}\left(\bm{\sigma}^*\right)/\sigma_{33}^*} \text{ and } A_{\parallel}^{\text{(*P)}} = A_{33}^*,
\end{align}	
which yields the desired conclusion.

\newpage
\subsection{$S$-tensor components}\label{Stenscomp}

\noindent
We now derive the second set of replacements under which the uniaxial bounds apply to fully anisotropic metamaterials. For this purpose, we will construct a second microstructure. As before, we denote the effective zmf conductivity tensor and the effective $S$-tensor of the fully anisotropic metamaterial as $\bm{\sigma}^*$ and $\bm{S}^*$, respectively. In contrast to the previous section, we initially assume that $\sigma_{12}^*=0$. This assumption will later be relaxed.\\

\noindent
First, we reflect the metamaterial at the $x_2x_3$-plane, yielding a mirrored version of the original metamaterial. The effective zmf conductivity tensor and the effective $S$-tensor of this mirrored version read
\begin{align}
	\bm{\sigma}^{\text{(*M)}}=\begin{pmatrix} \sigma_{11}^{*} && 0 && -\sigma_{13}^{*} \\
		0 && \sigma_{22}^{*} && \sigma_{23}^{*} \\
		-\sigma_{31}^{*} && \sigma_{32}^{*} && \sigma_{33}^{*} \end{pmatrix}
	\text{ and }
	\bm{S}^{\text{(*M)}}=\begin{pmatrix} S_{11}^{*} && -S_{12}^{*} && -S_{13}^{*} \\
		-S_{21}^{*} && S_{22}^{*} && S_{23}^{*} \\
		-S_{31}^{*} && S_{32}^{*} && S_{33}^{*} \end{pmatrix},
\end{align}
respectively. Next, we laminate the original metamaterial and its reflected version in equal proportions with layers perpendicular to the $x_1$-direction. Thereby, we obtain a laminate with the following effective properties
\begin{align}
	\bm{\sigma}^{\text{(*2)}}=\begin{pmatrix} \sigma_{11}^{*} && 0 && 0 \\
		0 && \sigma_{22}^{*} && \sigma_{23}^{*} \\
		0 && \sigma_{32}^{*} && \sigma_{33}^{*}-\frac{\left(\sigma_{13}^{*}\right)^2}{\sigma_{11}^{*}} \end{pmatrix} 
	\text{ and }
	\bm{S}^{\text{(*2)}}=\begin{pmatrix} S_{11}^{*}+\frac{\sigma_{13}^{*}}{\sigma_{11}^{*}}S_{31}^{*} && 0 && 0 \\
		0 && S_{22}^{*} && S_{23}^{*} \\
		0 && S_{32}^{*} && S_{33}^{*} \end{pmatrix}.
\end{align}	
We proceed with a second lamination step. Again we laminate the metamaterial obtained in the previous step with a copy of it that is mirrored at the $x_1x_3$-plane in equal proportions with layers perpendicular to the $x_1$-direction. The effective properties of the resulting laminate are
\begin{align}
	\bm{\sigma}^{\text{(*3)}}=\text{diag}\left(\sigma_{11}^{*}, \sigma_{22}^{*}, \sigma_{33}^{*}-\frac{\left(\sigma_{13}^{*}\right)^2}{\sigma_{11}^{*}}\right)
	\text{ and }
	\bm{S}^{\text{(*3)}}=\text{diag}\left(S_{11}^{*}+\frac{\sigma_{13}^{*}}{\sigma_{11}^{*}}S_{31}^{*}, S_{22}^{*}, S_{33}^{*}\right).
\end{align}	
As above, in the last step, we form a uniaxial polycrystal from this metamaterial. Specifically, we construct an extruded (columnar) version of a checkerboard. The squares of one of the two colors (e.g., the white squares) are made from the metamaterial formed in the previous step. In the squares of the other color (e.g., the black squares), the metamaterial is rotated by $90^{\circ}$ about the $x_3$-axis. This polycrystal has effective properties 
\begin{align}
	\sigma_{\perp}^{\text{(*P)}} = \sqrt{\sigma_{11}^*\sigma_{22}^*} \text{ and } S_{\parallel}^{\text{(*P)}} = S_{33}^*.
\end{align}

\noindent
To generalize beyond the original assumption $\sigma_{12}^*=0$, we perform a rotation about the $x_3$-axis that eliminates this component. In the original coordinate system (i.e., prior to the rotation), this yields
\begin{align}
	\sigma_{\perp}^{\text{(*P)}} = \sqrt{\sigma_{11}^*\sigma_{22}^*-(\sigma_{12}^*)^2}=\sqrt{\text{Cof}\left(\bm{\sigma}^{*}\right)_{33}} \text{ and } S_{\parallel}^{\text{(*P)}} = S_{33}^*.
\end{align}
thereby confirming the validity of the alternative replacements for applying the uniaxial bounds to fully anisotropic metamaterials.

\newpage
\section{Analytic expressions}

\noindent
In this section, we will provide analytic expressions for the volume-fraction dependent two-phase bounds in the effective parameter plane (Subsection~\ref{analytictwophase}) as well as for the bounds on the diagonal components of the effective Hall tensor as a function of the conductivity contrast (Subsection~\ref{Halltensbound}).

\subsection{Two-phase bounds}\label{analytictwophase}

\noindent
We consider uniaxial metamaterials formed from two isotropic phases with conductivities $\sigma_1$ and $\sigma_2<\sigma_1$ and $S$-coefficients $S_1>0$ and $S_2=0$. As discussed in Section 5.1 of the main text, the effective properties of such metamaterials are confined to the wedge in the $y$-plane defined by Equation\,(42), which we repeat here,
\begin{equation}
	|y_S| \leq \frac{y_{\sigma\perp}-\sigma_2}{\sigma_1-\sigma_2}|S_1B|. \nonumber
\end{equation}
Via the inverse $y$-transformation, we obtain corresponding expressions in the effective parameter plane, i.e., in the $\sigma_{\perp}^*$-$S_{\parallel}^*$-plane. In this plane, the effective axial $S$-coefficient is confined to a lens-shaped region defined by an upper and a lower bound,
\begin{equation}
	S_{\text{vf}}^{\text{lower}}(\sigma_{\perp}^*,f_1) \leq S_{\parallel}^* \leq S_{\text{vf}}^{\text{upper}}(\sigma_{\perp}^*,f_1),
	\label{eq:boundsintro}
\end{equation}
wherein the upper bound is given by 
\begin{equation}
	S_{\text{vf}}^{\text{upper}}(\sigma_{\perp}^*,f_1) = \frac{f_1(\sigma_1-\sigma_2)(\sigma_1+3\sigma_2)(\sigma_{\perp}^*-\sigma_2)-2\sigma_2(\sigma_2-\sigma_{\perp}^*)^2-f_1^2 (\sigma_1 - \sigma_2)^2 (\sigma_2 +\sigma_{\perp}^*)}{f_1f_2 (\sigma_1 - \sigma_2)^3}S_1
\end{equation}
and the lower bound reads
\begin{equation}
	S_{\text{vf}}^{\text{lower}}(\sigma_{\perp}^*,f_1) = \frac{f_1(\sigma_1-\sigma_2)(\sigma_1+\sigma_2+2\sigma_{\perp}^*)(\sigma_2-\sigma_{\perp}^*)+2\sigma_1(\sigma_2-\sigma_{\perp}^*)^2+f_1^2(\sigma_1-\sigma_2)^2(\sigma_2+\sigma_{\perp}^*)}{f_1f_2 (\sigma_1 - \sigma_2)^3}S_1.
\end{equation}
Note that this expression is significantly more complicated than the corresponding expression in the $y$-plane. In particular, the latter hides the volume-fraction dependence of the bounds.
For $S_1<0$, the upper and lower bound flip roles. Expressions for $S_2\neq0$ can be derived straightforwardly via a corresponding translation.

\subsection{Bounds on the diagonal Hall tensor components}\label{Halltensbound}

\noindent
We continue the discussion from Section 5.3 of the main text. Recall that we are again considering metamaterials formed from two isotropic phases with conductivities $\sigma_1$ and $\sigma_2<\sigma_1$ and Hall coefficients $A_1\neq 0$ and $A_2=0$. As discussed in the main text, the range of any diagonal effective Hall tensor component, $A_{ii}^*$, increases with increasing conductivity contrast of the phases, $\sigma_1/\sigma_2$. For low conductivity contrasts, the pure phases are optimal. Above $\sigma_1/\sigma_2=2$ enhancements and above $\sigma_1/\sigma_2=9$ sign-inversions are permitted by the bounds. More generally, $A_{ii}^*/A_1$ has to lie between an upper and a lower bound with the corresponding analytic expressions given by 
\begin{align}
	&A^{\text{upper}} = \left\{
	\begin{array}{ll}
		1 & \text{if }\sigma_1 \leq 2 \sigma_2 \\
		\frac{\sigma_1^2}{4\sigma_2(\sigma_1-\sigma_2)} & \text{if }\sigma_1 > 2 \sigma_2, \nonumber\\
	\end{array}\right. \\
	&A^{\text{lower}} = \left\{
	\begin{array}{ll}
		0 & \text{if }\sigma_1 \leq 9 \sigma_2 \\
		-\frac{\sigma_1^2(\sqrt{\sigma_1}-3\sqrt{\sigma_2})^2(\sigma_1-\sigma_2+2\sqrt{\sigma_1 \sigma_2})}{4\sigma_2(\sigma_1-\sigma_2)(\sigma_1^2-6\sigma_1 \sigma_2 + \sigma_2^2)} & \text{if }\sigma_1 > 9 \sigma_2. \\
	\end{array}\right.~~~~~
\end{align}
Note that, through the polycrystal argument discussed above, this result is not limited to uniaxial metamaterials. The upper bound is optimal for uniaxial metamaterials as it is attained by a rank-$1$ laminate. Results for $A_2\neq0$ can, in principle, be obtained via a suitable translation although this is more involved as it is the antisymmetric part of the conductivity tensor and not of the resistivity tensor that is being shifted (translations of the resistivity tensor do not have the required invariance properties).

\end{document}